\documentclass[twocolumn,showpacs,superscriptaddress,citeautoscript,byrevtex,bibnotes,footinbib,pre,aps]{revtex4}

\usepackage{graphicx}
\usepackage{amsmath}

\def\Jin{{\cal J}_{\text{in}}}
\def\Jout{{\cal J}_{\text{out}}}
\def\Jalpha{{\cal J}_{\alpha}}
\def\E{{\cal E}}

\def\rotor{\vec{\nabla}\times}
\def\r{\vec{r}}
\def\rr{\r\,'}
\def\freq{\left(\frac{\omega}{c}\right)^2}
\def\G{\hat{G}}
\def\Green{\G(\r,\rr|\omega)}
\def\eps{\hat{\varepsilon}}
\def\deps{\delta\eps}
\def\I{\hat{I}}

\begin{document}

\title{All-optical switching, bistability, and slow-light transmission \\
in photonic crystal waveguide-resonator structures}

\author{Sergei F. Mingaleev}

\affiliation{ Institut f\"ur Theoretische Festk\"orperphysik,
  Universit\"at Karlsruhe,  Karlsruhe 76128, Germany}

\affiliation{Bogolyubov Institute for Theoretical Physics of the
National Academy of Sciences of Ukraine, 03143 Kiev, Ukraine}

\author{Andrey E. Miroshnichenko}

\affiliation{Nonlinear Physics Centre and Centre for Ultra-high
bandwidth Devices for Optical Systems (CUDOS), Research School of
Physical Sciences and Engineering, Australian National University,
Canberra ACT 0200, Australia}

\author{Yuri S. Kivshar}

\affiliation{Nonlinear Physics Centre and Centre for Ultra-high
bandwidth Devices for Optical Systems (CUDOS), Research School of
Physical Sciences and Engineering, Australian National University,
Canberra ACT 0200, Australia}

\author{Kurt Busch}

\affiliation{ Institut f\"ur Theoretische Festk\"orperphysik,
Universit\"at Karlsruhe,  Karlsruhe 76128, Germany}

\begin{abstract}
We analyze the resonant linear and nonlinear transmission through a
photonic crystal waveguide side-coupled to a Kerr-nonlinear photonic
crystal resonator.
Firstly, we extend the standard coupled-mode theory analysis to photonic
crystal structures and obtain explicit analytical expressions for the
bistability thresholds and transmission coefficients which provide the
basis for a detailed understanding of the possibilities associated with
these structures.
Next, we discuss limitations of standard coupled-mode theory and present
an alternative analytical approach based on the effective discrete equations
derived using a Green's function method. We find that the {\em discrete
nature} of the photonic crystal waveguides allows a novel,
{\em geometry-driven} enhancement of nonlinear effects by shifting the
resonator location {\em relative} to the waveguide, thus providing an
additional control of resonant waveguide transmission and Fano resonances.
We further demonstrate that this enhancement may result in the lowering
of the bistability threshold and switching power of nonlinear devices
by several orders of magnitude. Finally, we show that employing such
enhancements is of paramount importance for the design of all-optical
devices based on {\em slow-light} photonic crystal waveguides.
\end{abstract}

\pacs{42.65.Pc; 42.70.Qs; 42.65.Hw; 42.79.Ta}
\date{May 17, 2006}

\maketitle

\section{Introduction}

It is believed that future integrated photonic circuits for ultrafast
all-optical signal processing require different types of nonlinear
functional elements such as switches, memory and logic devices.
Therefore, both novel physics and novel designs of such all-optical
devices have attracted significant research efforts during the last
two decades, and most of these studies utilize the concepts of optical
switching and bistability~\cite{Gibbs:1985:Book}.

One of the simplest bistable optical devices which can find applications
in photonic integrated circuits is a two-port device which is connected
to other parts of a circuit by one input and one output waveguide. Its
transmission properties depend on the intensity of light sent to the
input waveguide. Two basic realizations of such a device can be provided
by either direct or side-coupling between the input and output waveguides
to an optical resonator. In the first case,  we obtain a system with
{\em resonant transmission} in a narrow frequency range, while in the
second case, we obtain a system with {\em resonant reflection}.
Both systems may exhibit optical bistability when the resonator is made
of a Kerr nonlinear material. The resonant two-port systems of the first
type, with direct-coupled resonator,
can be realized in one-dimensional systems, and they have been
studied in great details in the context of different applications. In
contrast, the resonant systems of the second type, with side-coupled
resonators, can only be realized in higher-dimensional structures, and
their functionalities are not yet completely understood.

\begin{figure}[tb]
\centerline{\scalebox{.35}{\includegraphics[clip]{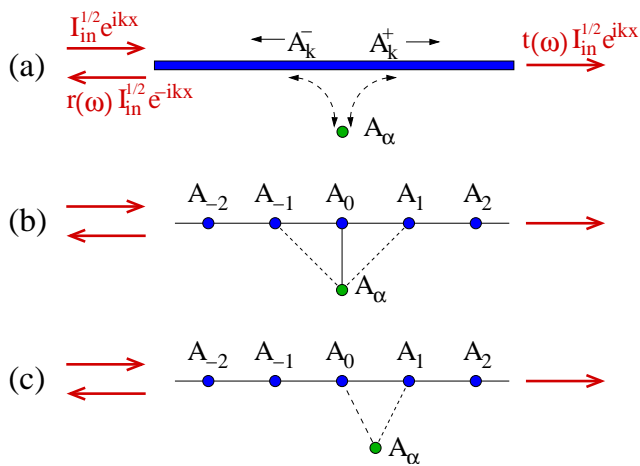}}}
\caption{(Color online) Three types of the geometries of a straight
photonic-crystal waveguide side-coupled to a nonlinear optical
resonator, $A_{\alpha}$. Standard coupled-mode theory is based on the
geometry (a) which does not account for discreteness-induced effects
in the photonic-crystal waveguides.
For instance, light transmission and bistability are qualitatively
different for (b) on-site and (c) inter-site locations of the resonator
along waveguide and this cannot be distinguished within the conceptual
framework of structure of type (a).}
\label{fig:struct}
\end{figure}

Our goal in this paper is to study in detail the second class of
resonant systems based on straight optical waveguides side-coupled
to resonators as shown in Fig.~\ref{fig:struct}.
Moreover, we assume that the waveguide and resonator are created
in two- or three-dimensional photonic crystal (PhC)~\cite{PhCs}.
Due to a periodic modulation of the refractive index of PhCs, such
structures may possess complete {\em photonic band gaps}, i.e.
regions of optical frequencies where PhCs act as ideal optical
insulators. Embedding carefully designed cavities into PhCs, one
can create ultra-compact {\em photonic crystal devices}
which are very promising for applications in photonic integrated
circuits. As an illustration, side-coupled waveguide-resonator
systems created in PhCs through arrays of cavities are schematically
depicted in Fig.~\ref{fig:struct}(b) and Fig.~\ref{fig:struct}(c).

Practical applications of such PhC devices are becoming a reality
due to the recent experimental success in realizing both linear and
nonlinear light transmission in two-dimensional PhC slab structures
where a lattice of cylindrical pores is etched into a planar
waveguide. In particular, Noda's group have realized coupling of a
PhC waveguide to a leaky resonator mode consisting of a defect
pore of slightly increased radius
\cite{Noda:2000-608:NAT,Chutinan:2001-2690:APL,Imada:2002-873:JLT,Asano:2003-407:APL}; Smith {\em et al.} demonstrated coupling of a three-line PhC
waveguide with a large-area hexagonal resonator
\cite{Smith:2001-1487:APL}; Seassal {\em et al.} have investigated
the mutual coupling of a PhC waveguide with a rectangular microresonator
\cite{Seassal:2002-811:IQE}; Notomi {\em et al.}
\cite{Notomi:2005-2678:OE} and Barclay {\em et al.}
\cite{Barclay:2005-801:OE} have observed all-optical bistability in
direct-coupled PhC waveguide-resonator systems.

Photonic-crystal based devices offer two major advantages over
corresponding ridge-waveguide systems: (i) the PhC waveguides may
have {\em very low group velocities} and, as a result, may
significantly enhance the effective coupling between short pulse
and resonators, and (ii) photonic crystals allow the creation of
ultra-compact high-Q resonators, which are essential for the
further miniaturization of all-optical nanophotonic devices.
Despite this, many researchers still believe that the basic
properties of devices based on ridge waveguides or PhC waveguides
are qualitatively identical, and that they can be correctly
described by the coupled-mode theory for continuous systems (see
Refs.~\cite{haus,Haus:1992-205:IQE,Fano:1961-1866:PREV,Anderson:1961-41:PREV,Fan:1998-960:PRL,Xu:2000-7389:PRE,Soljacic:2002-55601:PRE,Yanik:2003-2739:APL,Chak:2006-035105:PRE,Cowan:2003-46606:PRE,Cowan:2005-R41:SST}
and the discussion in Sec.~II).

However, an inspection of Figs.~\ref{fig:struct}(a-c) reveals, that
a major difference between the ridge waveguide in (a) and PhC
waveguides in (b,c) is that a PhC waveguide is always created by
an array of coupled small-volume cavities and, therefore, exhibits
an inherently {\em discrete nature}. This suggests that in these
systems an additional coupling parameter appears which relates
the position of the $\alpha$-resonator  to the waveguide cavities
along the waveguide. As a matter of fact, we may (laterally) place
the $\alpha$-resonator at any point relative to two succesive
waveguide cavities., thus creating a generally {\em asymmetric
device} which (in the nonlinear transmission regime) should exhibit
the properties of an optical diode, i.e., transmit high-intensity
light in one direction only. This is an intriguing peculiarity
of photonic-crystal based devices which we will analyze in a
future publication. In this paper, however, we restrict our analysis
to symmetric structures and study the cases of either {\em on-site
coupling} of the $\alpha$-resonator to the PhC waveguide, shown
schematically in Figs.~\ref{fig:struct}(b), or {\em inter-site
coupling}, as shown in Fig.~\ref{fig:struct}(c).

To address these issues, we employ a recently developed
approach~\cite{Miroshnichenko:2005-36626:PRE,Miroshnichenko:2005-56611:PRE,Miroshnichenko:2005-3969:OE}
and describe the photonic-crystal devices via effective discrete
equations that are derived by means of a Green's function
formalism~\cite{Mingaleev:2002-231:OL,Mingaleev:2002-2241:JOSB,McGurn:1999:PLA,Mingaleev:2000-5777:PRE,Mingaleev:2001-5474:PRL}.
This approach allows us to study the effect of the discrete nature
of the device on its transmission properties. In particular, we
show that the transmission depends on the location of the
resonance frequency $\omega_{\alpha}$ of the $\alpha$-resonator
with respect to the edges of the waveguide passing band. If
$\omega_{\alpha}$ lies deep inside the passing band, all devices
shown in Figs.~\ref{fig:struct}(a-c) are qualitatively similar,
and can adequately be described by the conventional coupled-mode
theory. However, if the resonator's frequency $\omega_{\alpha}$
moves closer to the edge of the passing band, standard
coupled-mode theory fails \cite{Waks:2005-5064:OE}. More
importantly, we show that in this latter case the properties of
the devices shown in Figs.~\ref{fig:struct}(b) and
Fig.~\ref{fig:struct}(c) become {\em qualitatively different}:
light transmission vanishes at both edges of the passing band, for
the cases shown in Fig.~\ref{fig:struct}(a) and
Fig.~\ref{fig:struct}(b), but for the case shown in
Fig.~\ref{fig:struct}(c) it remains perfect at one of the edges.
Moreover, the resonance quality factor for the structure (c) grows
indefinitely as $\omega_{\alpha}$ approaches this latter band
edge, accordingly reducing the threshold intensity required for a
bistable light transmission. This permits to achieve a very
efficient all-optical switching in the {\em slow-light} regime.

The paper is organized as follows. In Sec.~II we summarize and
extend the results of standard coupled-mode theory which
accurately describes the system shown in Fig.~\ref{fig:struct}(a).
Then, in Sec.~III.A we derive a system of effective discrete
equations \cite{Mingaleev:2002-231:OL,Mingaleev:2002-2241:JOSB}
and utilize a recently developed approach for its
analysis~\cite{Miroshnichenko:2005-36626:PRE,Miroshnichenko:2005-56611:PRE}.
Specifically, in Sec.~III.B and Sec.~III.C, respectively, we study
the two geometries of the waveguide-resonator coupled systems
schematically depicted in Fig.~\ref{fig:struct}(b) and
Fig.~\ref{fig:struct}(c). In Sec.~IV, we illustrate our main
findings for several examples of optical devices based on a
two-dimensional photonic crystal created by a square lattice of Si
rods. Finally, in Sec.~V we summarize and discuss our results. For
completeness as well as for justification of the effective
discrete equations employed, we include in Appendix~A an analysis
of simpler cases of uncoupled cavities and waveguides. The effects
of nonlocal waveguide dispersion and nonlocal waveguide-resonator
couplings are briefly summarized in Appendix~B.

\section{Coupled-mode theory}

In this Section, we first summarize the results of standard
coupled-mode theory and other similar approaches developed for
the analysis of continuous-waveguide structures similar to those
displayed in Fig.~\ref{fig:struct}(a). Then, we extend these
results in order to obtain {\em analytical formulas} for the
description of bistable nonlinear transmission in such devices.

\subsection{Linear transmission}

Transmission of light in waveguide-resonator systems is usually
studied in the linear limit using a coupled-mode theory based on a
Hamiltonian approach. This approach has been pioneered by Haus and
co-workers~\cite{haus,Haus:1992-205:IQE} and is similar to that
used by Fano~\cite{Fano:1961-1866:PREV} and
Anderson~\cite{Anderson:1961-41:PREV} for describing the
interaction between localized resonances and continuum states in
the context of an effect which is generally referred to as ``Fano
resonance''. For the analysis of the transmission of
photonic-crystal devices, this approach has been employed first by
Fan {\em et al.}~\cite{Fan:1998-960:PRL} and has been elaborated
on by Xu {\em et al.}~\cite{Xu:2000-7389:PRE}.

Throughout this paper we consider the propagation of a
monochromatic wave with the frequency $\omega$ lying inside the
waveguide passing band; we assume that the waveguide is
single-moded as well as that the resonator $\alpha$ is
non-degenerate and losses can be neglected. In this case, the
complex transmission and reflection amplitudes, 
$t(\omega)$ and $r(\omega)$, can be written in the form
\begin{eqnarray}\label{tr-Fano}
t(\omega)=\frac{\sigma(\omega)}{\sigma(\omega)-i},  \quad r(\omega)=
\frac{e^{i\varphi_r(\omega)}}{\sigma(\omega)-i},
\end{eqnarray}
with a certain real-valued and frequency-dependent function
$\sigma(\omega)$ and the reflection phase $\varphi_r(\omega)$.
Accordingly, the absolute values of the transmission coefficient
$T=|t|^2$ and reflection coefficient $R=|r|^2$ are
\begin{eqnarray}
\label{tr-Fano-abs}
T(\omega)= \frac{\sigma^{2}(\omega)}{\sigma^{2}(\omega) + 1}
\quad \text{and} \quad
R(\omega)= \frac{1}{\sigma^{2}(\omega)+1} \; ,
\end{eqnarray}
and it is easy to see that $T+R = 1$ for any $\sigma(\omega)$.

If the frequency $\omega_{\alpha}$ of the resonator $\alpha$ lies
inside the waveguide passing band, Fano-like resonant scattering
with zero transmission at the {\em resonance frequency}
$\omega_{\text{res}}$, lying in the vicinity of the resonator's
frequency, $\omega_{\alpha}$, should be
observed~\cite{Fano:1961-1866:PREV,Miroshnichenko:2005-36626:PRE}.
This corresponds to the condition $\sigma(\omega_{\text{res}})=0$
and, based on the terminology developed in
Refs.~\cite{Soljacic:2002-55601:PRE,Yanik:2003-2739:APL},
$\sigma(\omega)$ may be interpreted as the detuning of the
incident frequency from resonance.

The results of standard coupled-mode theory analysis (for
instance, see Ref.~\cite{Xu:2000-7389:PRE}) indicate that in the
vicinity of a high-quality (or high-Q) resonance, the detuning
function $\sigma(\omega)$ can be accurately described through the
linear function
\begin{eqnarray}\label{sigma-Haus}
\sigma(\omega) \simeq \frac{\omega_{\text{res}}-\omega}{\gamma} \; ,
\quad \text{where} \quad \gamma=\frac{\omega_{\text{res}}}{2Q} \; ,
\end{eqnarray}
which leads to a Lorentzian spectrum. Here, $Q$ is the quality
factor of the resonance mode of the $\alpha$-resonator. From the
Hamiltonian approach~\cite{Xu:2000-7389:PRE}, we find that the
resonance frequency $\omega_{\text{res}}$ almost coincides with
the resonator frequency $\omega_{\alpha}$ (see, however, Appendix
A in Ref.~\cite{Chak:2006-035105:PRE} for a more accurate estimate
of $\omega_{\text{res}}$), the reflection phase is $\varphi_r =
\pi/2$, and the resonance width $\gamma$ is determined by the
overlap of the mode profiles of waveguide and resonator:
\begin{eqnarray}\label{gamma-Haus}
\gamma \approx \frac{L}{v_{\text{gr}}}
\frac{\omega_{\text{res}}^2}{4 W_{k} W_{\alpha}} \, \left[ \int
d\vec{r} \, \delta\varepsilon(\vec{r}) \vec{\E}_{k}^{*}(\vec{r})
\vec{\E}_{\alpha}(\vec{r}) \right]^2 \; .
\end{eqnarray}
Here, $\vec{\E}_{\alpha}(\vec{r})$ is the normalized dimensionless
electric field of the resonator mode, $\vec{\E}_{k}(\vec{r})$ is
the corresponding field of the waveguide mode at wavevector
$k=k(\omega_{\text{res}})$, $v_{\text{gr}}=(d\omega/dk)$ is the
group velocity calculated at the resonance frequency, and $L$ is
the length of the waveguide section employed for the normalizing
the modes to
\begin{eqnarray}\label{E-norm}
\int\limits_{\text{wg section}} \!\!\!\!\!\!\! d\vec{r} \,
\varepsilon_{\text{wg}}(\vec{r}) |\vec{\E}_{k}(\vec{r})|^2
 = W_{k} \; , \nonumber \\ \int\limits_{\text{all
space}} \!\!\!\!\!\!\! d\vec{r} \, \varepsilon_{\alpha}(\vec{r})
|\vec{\E}_{\alpha}(\vec{r})|^2 = W_{\alpha} \; .
\end{eqnarray}
Furthermore, $\varepsilon_{\alpha}(\vec{r})$ and
$\varepsilon_{\text{wg}}(\vec{r})$ are the dielectric functions
that describe the resonator $\alpha$ and waveguide, respectively.
From Eqs.~(\ref{gamma-Haus})--(\ref{E-norm}) it is easy to see that
the resonance width $\gamma$ does not depend on the length $L$.

However, within the Hamiltonian approach, the function
$\delta\varepsilon(\vec{r})$ in Eq.~(\ref{gamma-Haus}) remains
undetermined. Generally, it is assumed to be a difference between
the total dielectric function and the dielectric function
$\varepsilon_0(\vec{r})$ ``associated with the unperturbed
Hamiltonian''~\cite{Xu:2000-7389:PRE} which is an ill-defined
quantity. A different approach based on a perturbative solution of
the wave equation for the electric
field~\cite{Cowan:2003-46606:PRE} sheds some light on the
resolution of this ambiguity and shows explicitly that
$\varepsilon_0(\vec{r})$ can be taken as either
$\varepsilon_{\text{wg}}(\vec{r})$ or
$\varepsilon_{\alpha}(\vec{r})$.

\subsection{Nonlinear transmission}
\label{sec:nonlin}

If the resonator $\alpha$ is made of a Kerr-nonlinear material,
increasing the intensity of the localized mode of the resonator
leads to a change of the refractive index and, accordingly, to a
shift of the resonator's resonance frequency. As a result, the
nonlinear light transmission in this case is described by the same
Eqs.~(\ref{tr-Fano})--(\ref{tr-Fano-abs}), with the only
difference that the frequency detuning parameter $\sigma(\omega)$
should be replaced by the generalized intensity-dependent
frequency detuning parameter $(\sigma(\omega)-\Jalpha)$. Here, 
$\Jalpha$ is a new dimensionless parameter which is, as we show
below, proportional to the intensity of the resonator's localized
mode. In particular, Eqs.~(\ref{tr-Fano-abs}) take the form
\begin{eqnarray}\label{tr-Fano-abs-nonlin}
T= \frac{[\sigma(\omega)-\Jalpha]^{2}} {[\sigma(\omega)-
\Jalpha]^{2} + 1} \; , \;\; R= \frac{1}{[\sigma(\omega)-
\Jalpha]^{2}+1} \; .
\end{eqnarray}
In order to find an explicit expression for $\Jalpha$, we assume 
that: ~(i) The dimensionless mode profiles
$\vec{\E}_{\alpha}(\vec{r})$ and $\vec{\E}_{k}(\vec{r})$
introduced in Eqs.~(\ref{gamma-Haus})--(\ref{E-norm}) are
normalized to their maximal values (as functions in real space),
i.e., $|\vec{\E}_{\alpha}(\vec{r})|^2_{\text{max}} =
|\vec{\E}_{k}(\vec{r})|^2_{\text{max}}=1$; ~(ii) The physical
electric fields are described by amplitudes, $A_{\alpha}$ and
$A_{k}$, multiplying the field profiles. Consequently, the maximum
intensity of the electric field in the vicinity of the
$\alpha$-resonator, $\vec{E}(\vec{r}) \simeq A_{\alpha}
\vec{\E}_{\alpha}(\vec{r})$, is equal to $|A_{\alpha}|^2$; ~(iii)
The $\alpha$-resonator is made of a Kerr-nonlinear material with
the nonlinear susceptibility $\chi^{(3)}_{\alpha}$ and it covers
the area described by the function $\theta_{\alpha}(\vec{r})$.
This function is equal to unity for all $\vec{r}$ inside the
cavities which form the resonator structure and vanishes outside.
In this case, $\Jalpha$ takes the form
\begin{eqnarray}\label{Jalpha}
\Jalpha = \frac{12\pi Q \, \kappa}{W_{\alpha}^2}
\chi^{(3)}_{\alpha} |A_{\alpha}|^2 \; ,
\end{eqnarray}
where $\kappa$ is the dimensionless and scale-invariant {\em
nonlinear feedback parameter} (first introduced in similar form in
Refs.~\cite{Soljacic:2002-55601:PRE,Yanik:2003-2739:APL}) which
measures the geometric nonlinear feedback of the system. It
depends on the overlap of the resonator's mode profile with
spatial distribution $\theta_{\alpha}(\vec{r})$ of nonlinear
material according to
\begin{eqnarray}\label{kappa}
\kappa = \frac{3}{W_{\alpha}^2} \left(
\frac{c}{\omega_{\text{res}}} \right)^d \!\!\!
\int\limits_{\text{all space}} \!\!\!\! d\vec{r} \,\,
\theta_{\alpha}(\vec{r}) \varepsilon_{\alpha}(\vec{r})
|\vec{\E}_{\alpha}(\vec{r})|^4 \; ,
\end{eqnarray}
where $d$ is the system dimensionality.

The dependence of $\Jalpha$ on the power of the incoming light has
already been studied analytically in
Refs.~\cite{Yanik:2003-2739:APL,Cowan:2003-46606:PRE,Cowan:2005-R41:SST}.
Here, we suggest a simpler form for this dependence
\begin{eqnarray}\label{Pin-Pa}
\Jin = \Jalpha \left( [\sigma(\omega)-\Jalpha]^{2} + 1 \right) \;
,
\end{eqnarray}
where we have introduced the dimensionless intensity $\Jin$ which
is proportional to the experimentally measured power of the
incoming light
\begin{eqnarray}\label{power-vs-Iin}
P_{\text{in}}= \frac{c^2 k(\omega)}{2\pi \omega} I_{\text{in}} =
P_{0} \Jin \; .
\end{eqnarray}
In this expression, we have abbreviated the incoming light
intensity as $I_{\text{in}}=|A_k|^2$ and introduced the {\em
characteristic power} $P_{0}$ of the waveguide defined as (see
Refs.~\cite{Yanik:2003-2739:APL,Soljacic:2002-55601:PRE,Cowan:2003-46606:PRE,Cowan:2005-R41:SST}
for derivation):
\begin{eqnarray}\label{P0}
P_{0} = \left(\frac{c}{\omega_{\text{res}}}\right)^{d-1}
\frac{\sqrt{\varepsilon_{\alpha}}}{Q^2 \, \kappa_{\alpha} \chi^{(3)}_{\alpha}} \; .
\end{eqnarray}
Finally, the outgoing light power $P_{\text{out}}=P_{0} \Jout$ can
be determined through the dimensionless intensity of the outgoing
light $\Jout = T\, \Jin$ with the transmission coefficient $T$
defined by Eq.~(\ref{tr-Fano-abs-nonlin}).

\begin{figure}[tbp]
\centerline{\scalebox{.35}{\includegraphics[clip]{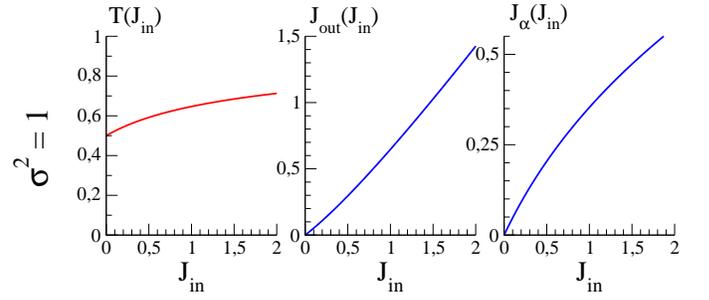}}}
\caption{(Color online) Dependencies of the transmission
coefficient, $T$, the outgoing light intensity, $\Jout$, and the
resonator's mode intensity, $\Jalpha$, on the incoming light
intensity, $\Jin$, for $\sigma^2(\omega)=1$ and negative product
$(\sigma(\omega) \cdot \Jalpha)$.}
\label{fig:T-L1A1-Iin-negative}
\end{figure}

It follows from Eqs.~(\ref{tr-Fano-abs-nonlin}) and (\ref{Pin-Pa})
that the nonlinear transmission problem is completely determined
by the value of $\sigma(\omega)$ and the sign of the product
$\sigma(\omega) \cdot \Jalpha$. As is illustrated in
Fig.~\ref{fig:T-L1A1-Iin-negative}, for frequencies where
$(\sigma(\omega) \cdot \Jalpha)<0$, the transmission coefficient
$T$ and the outgoing light intensity $\Jout$ grow monotonically
with $\Jin$ for all values of $\sigma(\omega)$.

\begin{figure}[tbp]
\centerline{\scalebox{.35}{\includegraphics[clip]{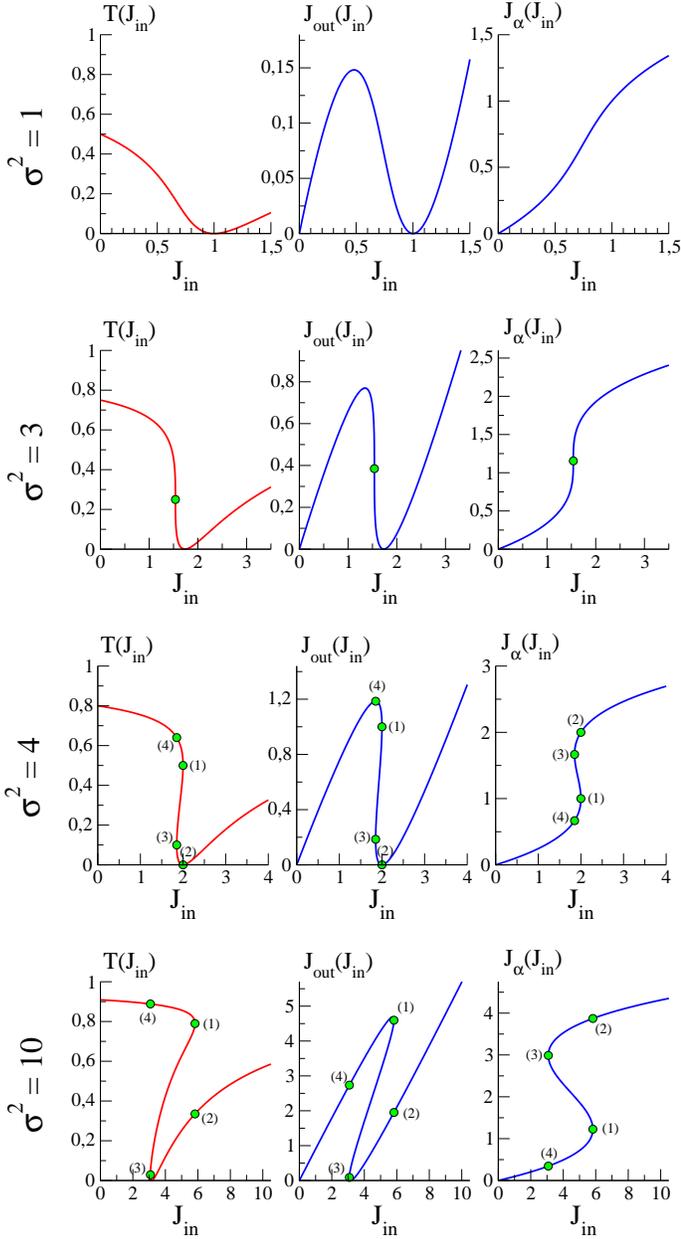}}}
\caption{(Color online) Dependencies of the transmission
coefficient, $T$, the outgoing light intensity, $\Jout$, and the
resonator's mode intensity, $\Jalpha$, on the incoming light
intensity, $\Jin$, for several different values of
$\sigma^2(\omega)$ and positive product $(\sigma(\omega) \cdot
\Jalpha)$.}
\label{fig:T-L1A1-Iin}
\end{figure}

\begin{figure}[tbp]
\centerline{\scalebox{.35}{\includegraphics[clip]{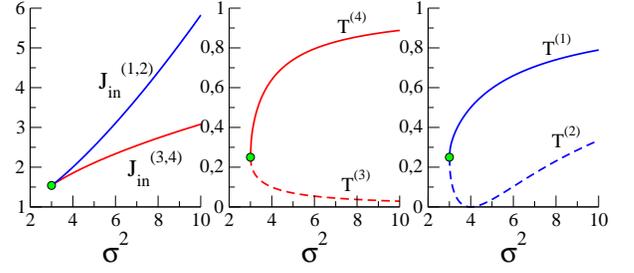}}}
\caption{(Color online) Dependencies of the threshold incoming
light intensity and the corresponding transmission coefficients on
$\sigma^2(\omega)$ for the four critical points (1)-(4) indicated
by circles in Fig.~\ref{fig:T-L1A1-Iin}. Here, we assume that
$(\sigma(\omega) \cdot \Jalpha)>0$.}
\label{fig:T-L1A1-sigma}
\end{figure}

The situation becomes more interesting for frequencies lying on the
other side of the resonance where $(\sigma(\omega) \cdot \Jalpha)>0$.
In this case $T$ and, therefore, $\Jout$ become
non-monotonic functions of $\Jin$, as is illustrated in
Fig.~\ref{fig:T-L1A1-Iin}. Moreover, for $\sigma^2(\omega) > 3$
these functions become {\em multi-valued} functions of $\Jin$ in the
interval $\Jin^{(3,4)} \leq \Jin \leq \Jin^{(1,2)}$, where
\begin{eqnarray}\label{IinMinMax}
\Jin^{(1,2)} = \frac{2}{27} \left\{ \sigma^3 + 9 \sigma + \left[
\sigma^2 - 3 \right]^{3/2} \right\} \; , \nonumber \\ \Jin^{(3,4)}
= \frac{2}{27} \left\{ \sigma^3 + 9 \sigma - \left[ \sigma^2 - 3
\right]^{3/2} \right\} \; ,
\end{eqnarray}
which are also shown in Fig.~\ref{fig:T-L1A1-sigma}. In this
interval the nonlinear light transmission becomes {\em bistable}:
low- and high-transmission regimes coexist at the same value of
the incoming light intensity $\Jin$, as can be seen in
Fig.~\ref{fig:T-L1A1-Iin} for $\sigma^2 > 3$ (intermediate parts
of the curves correspond to unstable transmission). Therefore, by
increasing an intially low intensity $\Jin$ we obtain a hysteris
where we jump from the point (1) to (2), and then upon decreasing
$\Jin$, we jump from the point (3) to (4). The transmission
coefficients at these characteristic points are
\begin{eqnarray}\label{TMinMax}
T^{(1,3)} &=& \frac{1}{2 \sigma^2
(1 \mp \sqrt{1-3/\sigma^2}) - 2} \; ,
\nonumber \\
T^{(2,4)} &=& \frac{(1 \mp 2\sqrt{1-3/\sigma^2})^2}
{5 - 3/\sigma^2 \mp 4 \sqrt{1-3/\sigma^2}} \; ,
\end{eqnarray}
and they are depicted in Fig.~\ref{fig:T-L1A1-sigma}. For
completeness, we also present the expressions for the resonator's
mode intensity at these points
\begin{eqnarray}\label{IalphaMinMax}
\Jalpha^{(1,3)} &=& \frac{2\sigma}{3} \mp \frac{1}{3}
\sqrt{\sigma^2-3} \; , \nonumber \\ \Jalpha^{(2,4)} &=&
\frac{2\sigma}{3} \pm \frac{2}{3} \sqrt{\sigma^2-3} \; .
\end{eqnarray}
From a practical point of view, these solutions have important
consequences. Firstly, the bistability condition $\sigma^2>3$
corresponds to a linear transmission $T>3/4$. That is, the
bistable transmission becomes possible only for frequencies where
$(\sigma(\omega) \cdot \Jalpha)$ is positive and linear
transmission exceeds 75\%. As demonstrated in
Fig.~\ref{fig:T-L1A1-sigma} and Eq.~(\ref{IinMinMax}), when
$\sigma^2$ grows, all threshold intensities grow, too, starting
with the minimum threshold intensity $\Jin^{(1,2,3,4)}=8/3^{1.5}
\approx 1.54$ at $\sigma^2=3$.

For ideal nonlinear switching the coefficients $T^{(1)}$ and
$T^{(4)}$ should be close to unity while $T^{(2)}$ and $T^{(3)}$
should vanish. However, as can be seen from
Fig.~\ref{fig:T-L1A1-sigma} and the asymptotic (for large
$\sigma^2$) expressions
\begin{eqnarray}\label{TMinMax-asym}
T^{(1)} &\approx& 1 - \frac{9}{\sigma^2} \; , \quad T^{(2)}
\approx 1 - \frac{9}{4 \sigma^2} \; , \nonumber \\ T^{(3)}
&\approx& 1 - \frac{1}{\sigma^2} \; , \quad T^{(4)} \approx
\frac{1}{4 \sigma^2} \; ,
\end{eqnarray}
of Eqs.~(\ref{TMinMax}), these conditions cannot be satisfied
simultaneously. In particular, the transmission coefficient
$T^{(2)}$ does not vanish but approaches unity for large
$\sigma^2$. Moreover, there exists no condition under which
$T^{(2)}$ and $T^{(3)}$ vanish simultaneously. Therefore, it is
impossible to create ideal nonlinear switches in these systems.

A reasonable compromise for realistic nonlinear switching schemes
of this type could be the usage of the frequency with $\sigma^2
\simeq 5$, for which the linear light transmission is close to
$83\%$. For this case, the critical transmission coefficients
$T^{(2)} \simeq 3.7\%$ and $T^{(3)} \simeq 7\%$ are sufficiently
small, while $T^{(1)} \simeq 60\%$ and $T^{(4)} \simeq 74\%$ are
large enough for practical purposes. The threshold intensities
$\Jin^{(1,2)} \simeq 2.53$ and $\Jin^{(3,4)} \simeq 2.11$ differ
about 20\% from each other, so that in this case one can achieve a
high-contrast and robust switching for sufficiently small
modulation of the incoming power.

The above analysis suggests that the optimal dimensionless
threshold intensities are fixed around $\Jin^{(i)} \sim 2.5$ so
that the real threshold power of the incoming light,
$P_{\text{in}}^{(i)} = P_{0}\Jin^{(i)}$, can only be minimized by
minimizing the characteristic power, $P_{0}$, of the system. An
inspection of Eq.~(\ref{P0}) shows that this can be facilitated by
increasing the resonator nonlinear feedback parameter,
$\kappa_{\alpha}$, the material nonlinearity,
$\chi^{(3)}_{\alpha}$, or the resonator quality factor, $Q$. For
small-volume photonic crystal resonators, it has been established
that $\kappa \sim 0.2$ (see
\cite{Yanik:2003-2739:APL,Soljacic:2002-55601:PRE}), and this
value can hardly be further increased.

Therefore, only two practical strategies remain that could lead to
an enhancement of nonlinear effects in this system. {\em The first
approach} is based on specific material properties: We should
create the resonator $\alpha$ from a material with {\em the
largest possible value} of $\chi^{(3)}_{\alpha}$. In high-index
semiconductors, nearly instantaneous Kerr nonlinearity reaches
values of $n_2 \sim 1.5 \times 10^{-13}$ cm$^2$/W
\cite{Sheik-Bahae:1991-1296:IQE}, where $n_2 \propto
\chi^{(3)}/n_0$ and $n_0$ is the linear refractive index of the
material. Even such relatively weak nonlinearity is already
sufficient for many experimental observations of the bistability
effect in the waveguide-resonator
systems~\cite{Notomi:2005-2678:OE,Barclay:2005-801:OE}. However,
using polymers with nearly instantaneous Kerr nonlinearity of the
order of $n_2 > 10^{-11}$ cm$^2$/W and, at the same time,
sufficiently weak two-photon absorption~\cite{Samoc:1995-1241:OL},
one could potentially decrease the value of $P_{0}$ by at least
two orders of magnitude. Polymers, however, have a low refractive
index which is insufficient for creating a (linear) photonic
bandgap required to obtain good waveguiding and low losses. The
solution to this could be the embedding of such highly nonlinear
but low-index materials into a host photonic crystal made of a
high-index semiconductor. Optimized waveguding designs for the
basic functional devices of this kind are available
~\cite{Mingaleev:2004-2858:OL,Schillinger:2005-324:SPIE,Jiao:2005-1875:IPTL}
and recent experimental progress
\cite{Heijden:2006-161112:APL,Ferrini:2006-1238:OL} may soon allow
a realization of corresponding linear and nonlinear devices.

{\em The second approach} is based on designing
waveguide-resonator structures with {\em the largest possible
quality factor}, $Q$. Potentially, one can increase $Q$
indefinitely by mere increase of the distance between the
waveguide and the resonator. However, this leads to a
corresponding increase in the size of the nonlinear photonic
devices. A very attractive alternative possibility for increasing
$Q$ is based on the adjustment of the resonator
geometry~\cite{Akahane:2003-944:NAT}.

In what follows, we suggest yet another possibility to
dramatically increase $Q$ through an optimal choice of the
resonator location relative to the discrete locations of the
cavities that form the photonic-crystal waveguide.

\subsection{Limitations of the coupled-mode theory}

Standard coupled-mode theory exhibits a number of limitations.
Firstly, it gives analytical expression for the detuning parameter
$\sigma(\omega)$ {\em only near the resonator frequency}
$\omega_{\alpha}$. And this immediately highlights the second
limitation: standard coupled-mode theory
\cite{Xu:2000-7389:PRE,Yanik:2003-2739:APL,Soljacic:2002-55601:PRE,Cowan:2003-46606:PRE,Cowan:2005-R41:SST,Chak:2006-035105:PRE}
cannot analytically describe resonant effects near waveguide band
edges. However, numerical studies~\cite{Waks:2005-5064:OE} have
recently demonstrated  that the effects of the waveguide
dispersion become very important at the band edges and may lead to
non-Lorentzian transmission spectra in coupled waveguide-resonator
systems.

As a matter of fact, the question ``what happens if the resonator
frequency $\omega_{\alpha}$ lies near the edge of the waveguide
passing band or even outside it?'' may be of a great practical
importance due to two reasons. Firstly, in realistic structures it
is not always possible to appropriately tune the frequency
$\omega_{\alpha}$, and therefore it is important to understand
properties of the system for any location of the resonance
frequency. Secondly, as we have already mentioned in the
Introduction, PhC waveguides can provide us with a
very slow group velocity of the propagating pulses --- but in most
cases they do it exactly at the passing band edges. Therefore, if we wish
to utilize such a slow light propagation for nonlinearity
enhancement, we should extend the above analysis to such cases,
too.

In what follows, we describe an alternative analytical approach to
the coupled waveguide-resonator structures which allows us to
correctly analyze both linear and nonlinear transmission for
arbitrary locations of the resonator frequency $\omega_{\alpha}$
relative to the waveguide passing band, including the transmission
near band edges in the slow light regime.

\section{Discrete model approach}

Having discussed the results obtained for the continuous-waveguide
structure shown in Fig.~\ref{fig:struct}(a), we now take into
account the discrete nature of the waveguding structure embedded
in photonic crystals. In particular, we analyze what will change
in the system properties when we move the resonator along the
waveguide from the on-site location shown in
Fig.~\ref{fig:struct}(b) to the inter-site location shown in
Fig.~\ref{fig:struct}(c). Our analysis is based on effective
discrete equations that have been derived for the description of
photonic crystal devices
\cite{McGurn:1999:PLA,Mingaleev:2000-5777:PRE,Mingaleev:2001-5474:PRL,Mingaleev:2002-2241:JOSB,Mingaleev:2002-231:OL}
in combination with a recently developed discrete model approach
to nonlinear Fano resonances~\cite{Miroshnichenko:2005-36626:PRE}.

\subsection{Discrete Equation Approach}
\label{sec:discrete-eqs}

First, we derive an appropriate set of discrete equations [see
Eqs.~(\ref{discrete-eqs-Fano}) below], and show that they can be
applied to a variety of the photonic-crystal devices. We start from
the wave equation in the frequency domain for the electric field
\begin{eqnarray}\label{Maxwell-eq}
\left[\rotor\rotor - \freq \eps(\r) \right]
\vec{E}(\r) = 0 \; ,
\end{eqnarray}
where the dielectric function
$\eps(\r)=\eps_{\text{pc}}(\r)+\deps(\r)$ consists of the
dielectric function $\eps_{\text{pc}}(\r)$ of a perfectly periodic
structure and a perturbation $\deps(\r)$ that describes the
embedded cavities. It is convenient to introduce the tensorial
Green function of the perfectly periodic photonic crystal,
\begin{eqnarray}\label{Green-eq}
\left[\rotor\rotor - \freq\eps_{\text{pc}}(\r) \right]
\Green = \I \delta(\r-\rr)
\end{eqnarray}
and to rewrite Eq.~(\ref{Maxwell-eq}) in the integral form,
\begin{eqnarray}\label{Green-eq-integral}
\vec{E}(\vec{r}) = \left( \frac{\omega}{c} \right)^2 \int
\vec{r}\,' \; \hat{G}(\vec{r},\vec{r}\,'|\omega) \,
\delta\hat{\varepsilon}(\vec{r}\,') \, \vec{E}(\vec{r}\,') \; ,
\end{eqnarray}
where we assume that the frequency $\omega$ lies inside a complete
photonic bandgap so that the electric field vanishes everywhere except
for areas inside and in the vicinity of cavities. We enumerate the
cavities by an integer index $n$ and introduce dimensionless functions
$\theta_n(\vec{r})$ which describe the shape of the $n$-th
cavity. As a result, $\deps(\r)$ may be represented as
\begin{eqnarray}\label{delta-epsilon}
\deps(\r) = \sum_n \left[
\deps_n + \chi^{(3)}_n |\vec{E}(\vec{r})|^2 \right]
\theta_n(\vec{r}-\vec{R}_n) \; ,
\end{eqnarray}
where $\vec{R}_n$, $\deps_n$, and $\chi^{(3)}_n$ are,
respectively, position, (linear) dielectric function, and
nonlinear third-order susceptibility of the $n$-th cavity.

Similar to Sec.~II, we describe the electric field of the $n$-th
cavity mode via a dimensionless field profile $\vec{\E}_n(\r)$ and
a complex amplitude $A_n$. Taking into account that {\em inside}
the cavities the electric field of the system is a superposition
\begin{eqnarray}\label{E-field-A}
\vec{E}(\vec{r}) \simeq \sum_n A_n \,
\vec{\E}_{n}(\vec{r}-\vec{R}_{n}) \; ,
\end{eqnarray}
Eq.~(\ref{Green-eq-integral}) can be rewritten as a set of
discrete nonlinear equations
\begin{eqnarray}\label{discrete-eqs}
D_{n}(\omega) A_n = \sum_{m \neq n} V_{n,m}(\omega) A_m +
\kappa_{n}(\omega) \chi^{(3)}_n |A_n|^2 A_n,
\end{eqnarray}
where $D_{n}(\omega)=1-V_{n,n}(\omega)$ is the dimensionless
frequency detuning from the resonance frequency, $\omega_n$,
of the $n$-th cavity. Furthermore,
\begin{eqnarray}\label{V-coeffs}
V_{n,m}(\omega) &=& \frac{\delta\varepsilon_m}{W_n}
\left(\frac{\omega}{c}\right)^2 \!\int\! d\vec{r} \!\int\!
d\vec{r}\,' \; \vec{\E}^{*}_n(\vec{r})
\hat{\varepsilon}_n(\vec{r})
\\ &\times& \theta_m(\vec{r}\,') \hat{G}(\vec{r}+\vec{R}_n-\vec{R}_m, \vec{r}\,' |
\omega) \, \vec{\E}_m(\vec{r}\,') \; , \nonumber
\end{eqnarray}
is the dimensionless linear coupling between the $n$-th and the $m$-th
cavity. Similarly,
\begin{eqnarray}\label{K-coeffs}
\kappa_{n}(\omega) &=& \frac{1}{W_n}
\left(\frac{\omega}{c}\right)^2 \!\int\! d\vec{r} \!\int\!
d\vec{r}\,' \; \vec{\E}^{*}_n(\vec{r})
\hat{\varepsilon}_n(\vec{r}) \\ &\times& \theta_n(\vec{r}\,')
\hat{G}(\vec{r}, \vec{r}\,' | \omega) \,
|\vec{\E}_n(\vec{r}\,')|^2 \vec{\E}_n(\vec{r}\,') \; , \nonumber
\end{eqnarray}
is the dimensionless and scale-invariant {\em nonlinear feedback
parameter} which should be compared with the analogous parameter
(\ref{kappa}) introduced in the conventional coupled-mode theory
analysis~\cite{Soljacic:2002-55601:PRE,Yanik:2003-2739:APL}.
Finally, $W_n$ is defined in exactly the same way as $W_{\alpha}$
in Eq.~(\ref{E-norm}).

We remark that in deriving Eqs.~(\ref{discrete-eqs}) we have
neglected higher-order couplings proportional to the integrals of
$\vec{\E}^{*}_n(\vec{r}) \vec{\E}_m(\vec{r}+\vec{R}_n-\vec{R}_m)$
with $n \neq m$ but take into account the coupling coefficients
which involve integrals of $\hat{G}(\vec{r}+\vec{R}_n-\vec{R}_m,
\vec{r}\,' | \omega)$ with $n \neq m$. This approximation is
sufficiently accurate in most cases, as we demonstrate in
Refs.~\cite{Mingaleev:2002-231:OL,Miroshnichenko:2005-3969:OE}. We
would like to mention that in
Eqs.~(\ref{discrete-eqs})--(\ref{K-coeffs}) we have used more
accurate definitions of the coupling coefficients than those that
have been introduced earlier in
Refs.~\cite{McGurn:1999:PLA,Mingaleev:2000-5777:PRE,Mingaleev:2001-5474:PRL,Mingaleev:2002-231:OL}.
They have also a more generic form than those we used in
Ref.~\cite{Mingaleev:2002-2241:JOSB}.

Typical frequency dependencies of the parameters of the discrete
model, Eq.~(\ref{discrete-eqs}), are displayed in
Figs.~\ref{fig:Dn}--\ref{fig:Vnm} of Appendix A, where we also
discuss the application of
Eqs.~(\ref{discrete-eqs})--(\ref{K-coeffs}) to simple structures
such as linear and nonlinear photonic crystal resonators and
straight waveguides. Here, we apply
Eqs.~(\ref{discrete-eqs})--(\ref{K-coeffs}) to study the more
complicated case of the nonlinear coupled waveguide-resonator
systems shown in Figs.~\ref{fig:struct}(b,c). The set of
Eqs.~(\ref{discrete-eqs}) may be separated in this case according
to
\begin{eqnarray}\label{discrete-eqs-Fano}
D_{\text{w}}(\omega) A_n &=& \sum_{j=1}^{L} V_{j\text{w}}(\omega)
(A_{n+j}+A_{n-j}) + V_{n,\alpha}(\omega) A_{\alpha} \; , \nonumber
\\ D_{\alpha}(\omega) A_{\alpha} &=& \sum_{j} V_{\alpha,j}(\omega)
A_{j} + \kappa_{\alpha}(\omega) \chi^{(3)}_{\alpha} |A_{\alpha}|^2
A_{\alpha} \; ,
\end{eqnarray}
where we assume that all cavities of the photonic-crystal
waveguide are identical and linear, so that we can denote
$D_{\text{w}}(\omega) \equiv D_{n}(\omega)$ and
$V_{j\text{w}}(\omega) \equiv V_{n,n\pm j}(\omega)$ for any $n$
inside the waveguide. Furthermore, the index $\alpha$ defines the
parameters of the side-coupled nonlinear resonator. Below we show
that the assumption of linear waveguide cavities may be relaxed
for frequencies near the resonator resonance frequency
$\omega_\alpha$ because then the amplitudes $A_n$ remain small in
comparison with the amplitude $A_{\alpha}$.

For the first equation in Eq.~(\ref{discrete-eqs-Fano}), we seek
solutions of standard form
\begin{eqnarray}\label{En-asymptotic}
A_n =
  \begin{cases}
    \; I_{\text{in}}^{1/2} \, t(\omega) \, e^{ik(\omega)sn}
         & \!\!\text{for $n \gg 1$}, \\
    \; I_{\text{in}}^{1/2} \, \Bigl[ e^{ik(\omega)sn} + r(\omega) \,
         e^{-ik(\omega)sn} \Bigr] &
         \!\!\text{for $n \ll -1$}, \\
  \end{cases}
\end{eqnarray}
where $s$ is the distance between the nearest waveguide cavities
and $I_{\text{in}}$ is the intensity of the incoming light. For
both structures shown in Figs.~\ref{fig:struct}(b,c), we obtain
that the transmission and reflection coefficients can formally be
described by the same expressions
(\ref{tr-Fano})--(\ref{tr-Fano-abs}) as for the structure depicted
in Fig.~\ref{fig:struct}(a). However, within the discrete equation
approach the expression for the detuning parameter
$\sigma(\omega)$ can now be found for the entire frequency range.
Below, we discuss novel results for the structures shown in
Fig.~\ref{fig:struct}(b) and Fig.~\ref{fig:struct}(c) separately.

\subsection{On-site resonator}
\label{sec:on-site}

First, we obtain the solution of this problem for the structure
shown in Fig.~\ref{fig:struct}(b). For simplicity, we assume that
the only nonvanishing coupling coefficients in
Eq.~(\ref{discrete-eqs-Fano}) are $V_{1\text{w}}(\omega)$,
$V_{\alpha,0}(\omega)$, and $V_{0,\alpha}(\omega)$ (see, however,
Appendix B for a more accurate analysis which takes into account
additional coupling coefficients). As a result, we obtain the
transmission and reflection coefficients described by
Eqs.~(\ref{tr-Fano})--(\ref{tr-Fano-abs}) with $\phi_{r}=\pi/2$
and a corresponding expression for $\sigma(\omega)$:
\begin{eqnarray}\label{sigma-local}
\sigma(\omega) = 2 \sin \left[ k(\omega) s \right] \,
\frac{V_{1\text{w}}(\omega)}{V_{0,\alpha}(\omega)}
\frac{A_0}{A_{\alpha}} \; ,
\end{eqnarray}
which should be considered as a generalized intensity-dependent
frequency detuning parameter $\sigma(\omega)+\Jalpha$ introduced
in Eq.~(\ref{tr-Fano-abs-nonlin}) above. The amplitude $A_0$ in
Eq.~(\ref{sigma-local}) is given by
\begin{eqnarray}\label{A0-local}
A_{0} &=& t(\omega) \, I_{\text{in}}^{1/2} \; ,
\end{eqnarray}
while the waveguide dispersion relation $k(\omega)$ is determined
by Eq.~(\ref{k-wave-TB}).

In the case of a linear $\alpha$-resonator (i.e.
$\chi^{(3)}_{\alpha} \equiv 0$), the amplitude $A_{\alpha} =
V_{\alpha,0}(\omega) A_{0}/D_{\alpha}(\omega)$ is proportional to
the amplitude $A_0$. Therefore, $\sigma(\omega)$ and, accordingly,
the transmission and reflection coefficients do not depend on the
light intensity. Upon introducing the abbreviation
\begin{eqnarray}\label{mu}
\mu(\omega) =
\frac{D_{\alpha}(\omega) V_{1\text{w}}(\omega)}
{V_{0,\alpha}(\omega) V_{\alpha,0}(\omega)} \; ,
\end{eqnarray}
the detuning parameter, Eq.~(\ref{sigma-local}), for a linear
$\alpha$-resonator reads as
\begin{eqnarray}\label{sigma-local-linear}
\sigma(\omega) = 2 \mu(\omega) \sin \left[ k(\omega)s \right] \; .
\end{eqnarray}
This implies that $\sigma(\omega)$ vanishes when either
$D_{\alpha}(\omega)=0$ or $k(\omega)=\pi n/s$ with an arbitrary
integer $n$. The first condition reproduces Eq.~(\ref{sigma-Haus})
with $\omega_{\text{res}} = \omega_{\alpha}$ and the resonance
width $\gamma$ given by
\begin{eqnarray}\label{gamma-local-model}
\gamma \approx \frac{\omega_{\alpha} \Delta_{\alpha}}
{\sin[k(\omega_{\alpha})s]} \approx \frac{s \,
\omega_{\alpha}\omega_{\text{w}}} {v_{\text{gr}}} \,  \nu_{\alpha}
\, \nu_{\text{w}} \, V_{0,\alpha} V_{\alpha,0} \; ,
\end{eqnarray}
where $\nu_{\alpha}$ and $\nu_{\text{w}}$ are defined by
Eq.~(\ref{Taylor-series-for-D}),
\begin{eqnarray}\label{Delta-B}
\Delta_{\alpha}=\frac{V_{0,\alpha} V_{\alpha,0}}{2
\omega_{\alpha}  D'_{\alpha} V_{1\text{w}}} =
\frac{V_{0,\alpha} V_{\alpha,0}}{2 V_{1\text{w}}}
\,  \nu_{\alpha} \; ,
\end{eqnarray}
and the group velocity
\begin{eqnarray}\label{v-gr}
v_{\text{gr}} = \left. \frac{d\omega}{dk} \right|_{\omega_{\alpha}}
\approx - 2 s \, \omega_{\text{w}} \nu_{\text{w}}
V_{1\text{w}} \sin [k(\omega_{\alpha}) s] \; ,
\end{eqnarray}
can be found directly from Eq.~(\ref{k-wave-TB}). Here and in what
follows, we assume that the values of all frequency-dependent
parameters whithout explicitly stated frequency dependence are
evaluated at the resonance frequency, $\omega_{\text{res}}$.
Finally, we notice that the resonance
width, Eq.~(\ref{gamma-local-model}), is very similar to that described
by the coupled-mode theory, Eq.~(\ref{gamma-Haus}).

\begin{figure}[tbp]
\centerline{\scalebox{.43}{\includegraphics[clip]{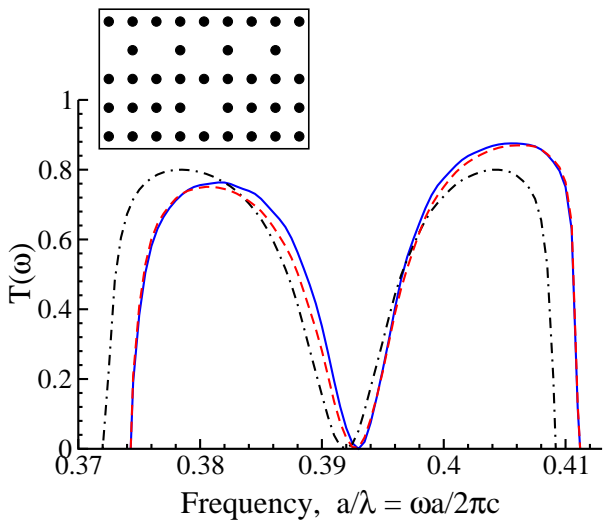}}}
\caption{(Color online) Linear transmission through a photonic crystal
waveguide that is created by removing every second rod in a row
($\vec{s}=2\vec{a}_1$) side-coupled to a one-site resonator created
by removing a single rod. The underlying 2D photonic crystal
is described in Appendix~\ref{sec:example-PhC}. We compare exact
numerical results (solid line) with the analytical results based
on Eq.~(\ref{sigma-local-linear}) (dot-dashed line) and
Eq.~(\ref{sigma-linear-L2}) (dashed line).}
\label{fig:transm-crow1-fano2}
\end{figure}

It is important that the quality factor $Q$ of the resonance
\begin{eqnarray}\label{Q-factor}
Q = \frac{\omega_{\alpha}}{2\gamma} \approx
\frac{\sin \left[k(\omega_{\alpha}) s \right]}{2\Delta_{\alpha}}
\approx \frac{v_{\text{gr}}}{s\, \omega_{\text{w}}} \,
\frac{1}{2 \, \nu_{\alpha} \nu_{\text{w}} \, V_{0,\alpha} V_{\alpha,0}} \; ,
\end{eqnarray}
is multiplied by the factor $\sin \left[ k(\omega_{\alpha}) s
\right] \sim v_{\text{gr}}$, and, therefore, becomes strongly
suppressed near the edges of waveguide passing band,
$k(\omega_{\alpha})=0, \, \pm\pi/s$. Accordingly, the detuning
parameter (\ref{sigma-local-linear}) vanishes at these edges, too.
This means that, in agreement with the numerical calculations
shown in Fig.~\ref{fig:transm-crow1-fano2}, the transmission
coefficient $T(\omega)$ vanishes not only at the resonance
frequency, but also at {\em both edges} of the waveguide passing
band. Such an effect was recently observed by Waks and
Vukovic~\cite{Waks:2005-5064:OE} in their numerical calculations
based on standard coupled-mode theory which takes into account the
waveguide dispersion. Therefore, the effect of vanishing
transmission at the spectral band edges may be attributed also to
the structure shown in Fig.~\ref{fig:struct}(a).

Obviously, this enhancement of light scattering at the waveguide
band edges should be very important from the point of view of
fabrication tolerances since virtually any imperfection
contributes to scattering losses. Moreover, as discussed in
Sec.~\ref{sec:discussion}, this effect is detrimental to the
concept of all-optical switching devices based on {\em slow-light}
photonic crystal waveguides.

We support this conclusion by another observation. First, the
light intensity at the $0$-th cavity, $|A_0|^2=T(\omega)
I_{\text{in}}$, vanishes at the resonance frequency for arbitrary
large incoming light intensity, because $T(\omega_{\alpha})\equiv
0$. Therefore, the nonlinearity of this cavity may safely be
neglected. In contrast, the light intensity at the
$\alpha$-resonator reaches its maximum value at $\omega_{\alpha}$,
\begin{eqnarray}\label{A-alpha}
|A_{\alpha}(\omega_{\alpha})|^2 \simeq 4 \left(
\frac{V_{1\text{w}}}{V_{0,\alpha}} \right)^2 \sin^2
[k(\omega_{\alpha})s] \cdot I_{\text{in}} \nonumber \\ \simeq
\left( \frac{v_{\text{gr}}}{s \omega_{\text{w}} \, \nu_{\text{w}}
V_{0,\alpha}} \right)^2  \cdot I_{\text{in}} \simeq \left( 2 Q \,
\nu_{\alpha} V_{\alpha,0} \right)^2 \cdot I_{\text{in}} \; ,
\end{eqnarray}
which may significantly exceed the incoming light intensity
$I_{\text{in}}$ when the coupling $V_{0,\alpha}$ between the
$\alpha$-resonator and waveguide becomes small enough relative to
the coupling $V_{1\text{w}}$ between the cavities in the
waveguide. This strong enhancement suggests a physical explanation
for the existence of the rather strong nonlinear effect of light
bistability at relatively low intensities of the incoming light.
However, when the resonance frequency $\omega_{\alpha}$ lies close
to any of the waveguide band edges, it is seen from
Eq.~(\ref{A-alpha}) that the light intensity at the
$\alpha$-resonator becomes (strongly) suppressed by a factor
$\sin^2 [k(\omega_{\alpha})s]$ .


Details of an extension of the above discussion to the case of
more realistic non-local couplings, i.e., more than nearest
neighbors couplings, is presented in Appendix B and here we only
summarize the results. Both, a non-locality of the inter-coupling
between waveguide cavities as well as a nonlocality of
cross-coupling with the $\alpha$-resonator lead to a small shift
in the resonance frequency, $\omega_{\text{res}}$, but do  not
change the main result about the suppression of the detuning
$\sigma(\omega)$ and transmission $T(\omega)$ at both edges of the
waveguide passing band. However, we would like to emphasize that
for a fully quantitative analysis, non-local couplings have to be
taken into account, for instance, within the framework of the
recently developed Wannier function approach
\cite{Busch:2003-R1233:JPCM}.

We now consider the case when the resonator $\alpha$ is nonlinear,
i.e. $\chi^{(3)}_{\alpha} \neq 0$. As has been previously shown in
Ref.~\cite{Miroshnichenko:2005-36626:PRE}, this case, too, can be
studied analytically even for non-local couplings between the
cavities and resonator and novel effects originating solely from
the non-locality may be expected when the non-local coupling
strength exceeds one half of the local coupling. Unfortunately, in
realistic photonic crystals this limit may hardly be realized so
that here we restrict our analysis to the local-coupling
approximation. In this case, we obtain from the second equation in
Eqs.~(\ref{discrete-eqs-Fano}) that the amplitude $A_{\alpha}$
uniquely determines the amplitude $A_{0}$. Substituting the latter
expression into Eqs.~(\ref{sigma-local})--(\ref{A0-local}), we
find that the nonlinear transmission is described by
Eqs.~(\ref{tr-Fano-abs-nonlin}) and (\ref{Pin-Pa}) with the
detuning $\sigma(\omega)$ determined by
Eqs.~(\ref{mu})--(\ref{sigma-local-linear}) and the dimensionless
intensities $\Jalpha$ and $\Jin$ given by the expressions
\begin{eqnarray}\label{Iin-Ia}
\Jalpha &\simeq& 2Q \kappa_{\alpha} \nu_{\alpha}
\chi^{(3)}_{\alpha} |A_{\alpha}|^2 \; ,
\\ \Jin &\simeq& 8
\sin[k(\omega_{\text{res}})s] \, V_{1\text{w}} \left(
\frac{\delta\varepsilon_{0}}{\delta\varepsilon_{\alpha}} \right)
Q^2 \kappa_{\alpha} \nu_{\alpha}^2 \chi^{(3)}_{\alpha}
I_{\text{in}} \nonumber \\ &\simeq& - \frac{4v_{\text{gr}}}{s \,
\omega_{\text{w}}} \left( \frac{\delta\varepsilon_{0} \nu_{\alpha}
}{\delta\varepsilon_{\alpha} \nu_{\text{w}}} \right) Q^2
\kappa_{\alpha} \nu_{\alpha} \chi^{(3)}_{\alpha} I_{\text{in}} \;
, \nonumber
\end{eqnarray}
where $Q$ is determined by Eq.~(\ref{Q-factor}). Therefore, all
the results for the nonlinear light transmission which are
displayed in
Figs.~\ref{fig:T-L1A1-Iin-negative}--\ref{fig:T-L1A1-sigma} are
directly applicable to the structure of Fig.~\ref{fig:struct}(b),
too.

In an experiment, one measures not the light intensity in the
waveguide, $I_{\text{in}}$, but the propagation power,
Eq.~(\ref{power-vs-Iin}), where for the discrete structure of
Fig.~\ref{fig:struct}(b), the characteristic power $P_{0}$ is
\begin{eqnarray}\label{P0-case1}
P_{0} &\simeq& \frac{c^2 k(\omega_{\alpha})}{16\pi
\sin[k(\omega_{\alpha})s] \omega_{\alpha} V_{1\text{w}}} \left(
\frac{\delta\varepsilon_{\alpha}}{\delta\varepsilon_{0}} \right)
\frac{1}{Q^2 \kappa_{\alpha} \nu_{\alpha}^2 \chi^{(3)}_{\alpha}}
\nonumber \\ &\simeq& - \frac{c^2 k(\omega_{\alpha}) s}{8\pi
v_{\text{gr}}} \left( \frac{\omega_{\text{w}}
\delta\varepsilon_{\alpha} \nu_{\text{w}}}{\omega_{\alpha}
\delta\varepsilon_{0} \nu_{\alpha}} \right) \frac{1}{Q^2
\kappa_{\alpha} \nu_{\alpha} \chi^{(3)}_{\alpha}} \; .
\end{eqnarray}
Again, this result is quite similar to Eq.~(\ref{P0}) for the
continuous structure of Fig.~\ref{fig:struct}(a). Nevertheless,
our more general analysis explicitly suggests that it should be
better to use the $\alpha$-resonator with the resonance frequency
at the center of the waveguide passing band $k(\omega_{\alpha})
\approx \pi/2s$, where the group velocity reaches its maximum.
Notice, however, that this suggestion becomes wrong for the
structure of Fig.~\ref{fig:struct}(c) studied in the next
subsection.

\subsection{Inter-site resonator}

In the system where the $\alpha$-resonator is placed {\em
symmetrically} between two cavities of the waveguide and,
therefore, couples equally to both of them, a qualitatively
different type of resonant transmission occurs. The corresponding
structure is schematically shown in Fig.~\ref{fig:struct}(c).
Assuming that in this case the nonvanishing coupling coefficients
in Eq.~(\ref{discrete-eqs-Fano}) are $V_{1\text{w}}(\omega)$,
$V_{\alpha,1}(\omega) \equiv V_{\alpha,0}(\omega)$, and
$V_{1,\alpha}(\omega) \equiv V_{0,\alpha}(\omega)$, we seek
solutions to the first equation of the
system~(\ref{discrete-eqs-Fano}) that are of the form of
Eq.~(\ref{En-asymptotic}). Again, we find that the transmission
and reflection coefficients are given by
Eqs.~(\ref{tr-Fano})--(\ref{tr-Fano-abs}) albeit with the
frequency-dependent phase $\phi_{r}(\omega)=\pi/2+k(\omega)s$.
Here, $k(\omega)$ is determined by Eq.~(\ref{k-wave-TB}), and the
generalized intensity-dependent frequency detuning is
\begin{eqnarray}\label{sigma-local-inter}
\sigma(\omega)+\Jalpha = i - i \left(e^{ik(\omega)s}-1\right)
\frac{V_{1\text{w}}(\omega)} {V_{0,\alpha}(\omega)}
\frac{I_{\text{in}}^{1/2}}{A_{\alpha}} \; .
\end{eqnarray}
The corresponding amplitudes are
\begin{eqnarray}\label{A0-local-inter}
A_{0} &=& I_{\text{in}}^{1/2} - \frac{1}{[1-e^{-ik(\omega)s}]}
\frac{V_{0,\alpha}(\omega)}{V_{1\text{w}}(\omega)} A_{\alpha} \; ,
\nonumber \\ A_{1} &=& e^{ik(\omega)s} I_{\text{in}}^{1/2} -
\frac{1}{[1-e^{-ik(\omega)s}]}
\frac{V_{0,\alpha}(\omega)}{V_{1\text{w}}(\omega)} A_{\alpha} \; .
\end{eqnarray}
Despite the complex form of Eq.~(\ref{sigma-local-inter}), we
would like to emphasize that the detuning $\sigma(\omega)$
determined by Eq.~(\ref{sigma-local-inter}) is a {\em real-valued}
function (see also the discussion following Eq.~(\ref{tr-Fano})
above).

\begin{figure}[tbp]
\centerline{\scalebox{.43}{\includegraphics[clip]{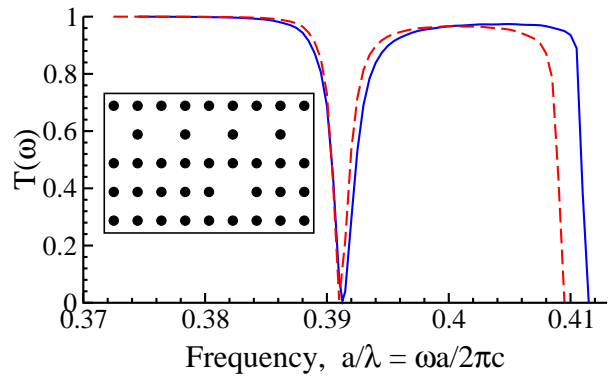}}}
\caption{(Color online) Liner transmission through a photonic
crystal waveguide created by removing every second rod in a row
($\vec{s}=2\vec{a}_1$) side-coupled to an inter-site resonator
created by removing a single rod. The underlying 2D photonic
crystal described in Appendix~\ref{sec:example-PhC}. We compare
exact numerical results (solid line) with the analytical results
based on Eq.~(\ref{sigma-local-linear-inter}) (dashed line).}
\label{fig:transm-crow1-fano2-inter}
\end{figure}

In the case of the linear $\alpha$-resonator (i.e., for
$\chi^{(3)}_{\alpha} \equiv 0$), we obtain
\begin{eqnarray}\label{sigma-local-linear-inter}
\sigma(\omega) &=& \left[ 1 + \mu(\omega) \right] \tan \left(
\frac{k(\omega)s}{2} \right) \; ,
\end{eqnarray}
where $\mu(\omega)$ is given by Eq.~(\ref{mu}). For a high-quality
$\alpha$-resonator in the vicinity of the resonance frequency this
detuning parameter can be approximated by Eq.~(\ref{sigma-Haus})
with
$\omega_{\text{res}} \simeq \omega_{\alpha} \left( 1-2\Delta_{\alpha} \right)$
and
\begin{eqnarray}\label{gamma-local-linear-inter}
\gamma &=& \frac{2 \omega_{\text{res}} \Delta_{\alpha}}{\tan[
k(\omega_{\text{res}})s/2]} \; .
\end{eqnarray}
Here, $\Delta_{\alpha}$ is defined by Eq.~(\ref{Delta-B}). 
In contrast to Eq.~(\ref{Q-factor}), the
corresponding quality factor
\begin{eqnarray}\label{Q-factor-inter}
Q = \frac{\omega_{\text{res}}}{2\gamma} \approx \frac{\tan
\left[k(\omega_{\text{res}}) s/2 \right]}{4\Delta_{\alpha}}
\approx \frac{V_{1\text{w}} \tan\left[k(\omega_{\text{res}}) s/2
\right]}{2 \, \nu_{\alpha} \, V_{0,\alpha} V_{\alpha,0}}
\end{eqnarray}
is now multiplied by the
factor $\tan[ks/2]$ which does not vanish and even diverges as
$k(\omega_{\text{res}})$ approaches the edge of the transmission
band $k=\pm\pi/s$. At this band edge, $\sigma(\omega) \sim
\tan(k(\omega)s/2) \to \infty$ and, therefore, light transmission
is always perfect. This conclusion is supported by the exact
numerical calculations presented in
Fig.~\ref{fig:transm-crow1-fano2-inter}. At the other band edge,
i.e., for $k(\omega)=0$, transmission vanishes, similar to the
structures shown in Figs.~\ref{fig:struct}(a,b).

The light intensity at the $\alpha$-resonator reaches its maximal
value at the resonance frequency
\begin{eqnarray}\label{A-alpha-inter}
|A_{\alpha}(\omega_{\text{res}})|^2 &\simeq& 4 \left(
\frac{V_{1\text{w}}}{V_{0,\alpha}} \right)^2 \sin^2
\left[\frac{k(\omega_{\text{res}})s}{2} \right] \cdot
I_{\text{in}} \nonumber \\ &\simeq& \left(4Q \nu_{\alpha}
V_{\alpha,0} \cos \left[\frac{k(\omega_{\text{res}})s}{2} \right]
\right)^2 \cdot I_{\text{in}} \;
\end{eqnarray}
Again, in contrast to the corresponding light intensity
(\ref{A-alpha}) for the on-site coupled structure,
Eq.~(\ref{A-alpha-inter}) does not vanish at the edge of the
transmission band $k=\pm\pi/s$. Therefore, we can expect that for
inter-site coupled structure nonlinear effects at the band edge
$k=\pm\pi/s$ should be sufficiently strong to allow bistable
transmission and switching.

To investigate this, we assume that the $\alpha$-resonator is
nonlinear ($\chi^{(3)}_{\alpha} \neq 0$) and introduce the same
dimensionless intensities $\Jalpha$ and $\Jin$ as in
Eq.~(\ref{Iin-Ia}). However, now the quality factor $Q$ is defined
by Eq.~(\ref{Q-factor-inter}) and the resonance frequency is
$\omega_{\text{res}} \simeq \omega_{\alpha}(1-2\Delta_{\alpha})$.
We find that this nonlinear problem, too, has a solution of the
form given by Eqs.~(\ref{tr-Fano-abs-nonlin}) and (\ref{Pin-Pa}).
However, now the detuning $\sigma(\omega)$ is given by
Eq.~(\ref{sigma-local-linear-inter}). Therefore, all results
presented above in
Figs.~\ref{fig:T-L1A1-Iin-negative}--\ref{fig:T-L1A1-sigma} remain
applicable to this structure, too. The only but very important
qualitative difference of the structure shown in
Fig.~\ref{fig:struct}(c) is that the transmission coefficient
$T(\omega)$ and the corresponding light intensity $|A_{\alpha}|^2$
at the $\alpha$-resonator {\em do not vanish} at the band edge
$k=\pm\pi/s$ since the quality factor $Q$ at this band edge grows
to infinity for the inter-site structure of
Fig.~\ref{fig:struct}(c). Therefore, this structure may be
utilized for realizing efficient all-optical switching devices
based on {\em slow-light photonic crystal waveguides}. This is in
sharp contrast to the structures shown in
Figs.~\ref{fig:struct}(a,b).

\section{Discussion of results}
\label{sec:discussion}

In this section, we summarize our results and emphasize
their importance by applying them to specific photonic-crystal
structures. We consider a two-dimensional photonic crystal created
by a square lattice of dielectric rods in air. The rods are made
from $Si$ or $GaAs$ ($\varepsilon=11.56$) and have radius
$r=0.18a$.

\begin{figure}[tbp]
\centerline{\scalebox{.43}{\includegraphics[clip]{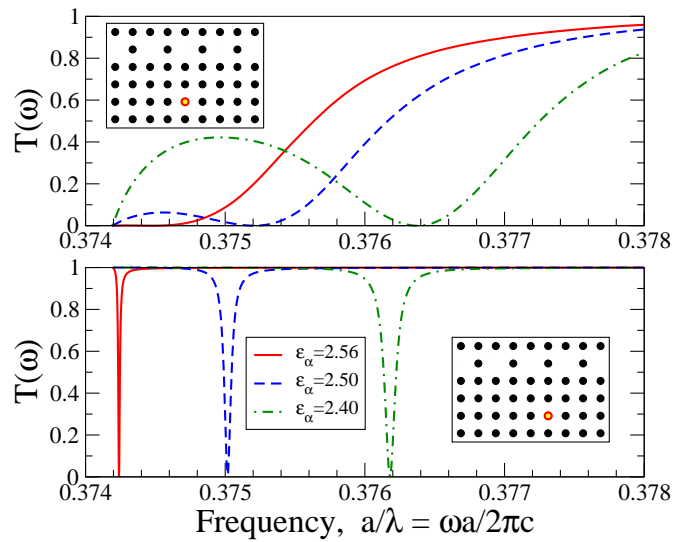}}}
\caption{(Color online) Linear transmission spectrum for a photonic
crystal waveguide created by removing every second rod in a row
($\vec{s}=2\vec{a}_1$) side-coupled to a single on-site
(a) or inter-site (b) polymer-rod resonator (marked by the open
circle in the insets). The underlying 2D photonic crystal is described
in Appendix~\ref{sec:example-PhC} and results for three different values
of the resonator dielectric constant $\varepsilon_{\alpha}$ are shown.}
\label{fig:example-air-wg}
\end{figure}

\begin{figure}[tbp]
\centerline{\scalebox{.43}{\includegraphics[clip]{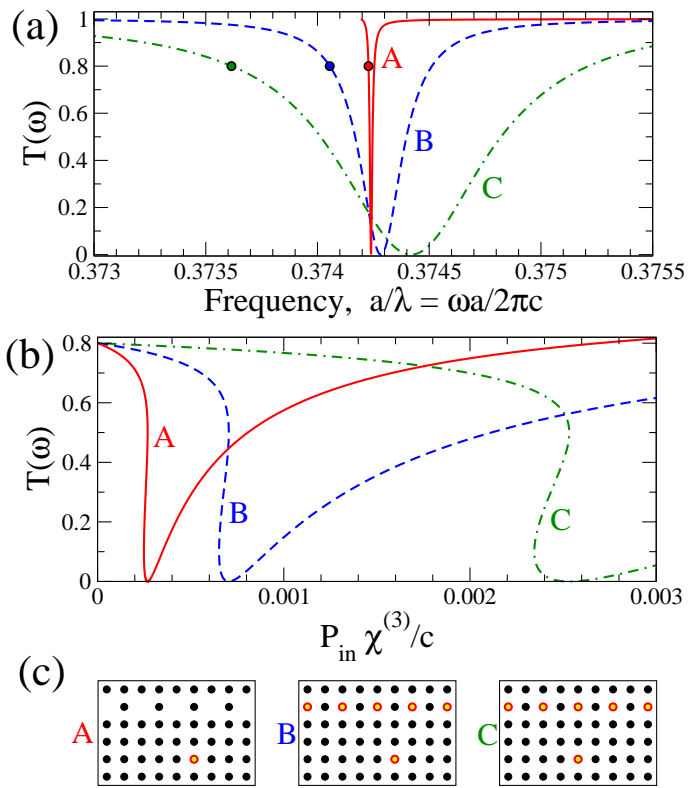}}}
\caption{(Color online) (a) Linear transmission spectrum and (b)
nonlinear bistable transmission for three different side-coupled
waveguide-resonator photonic crystal structures whose topology
is shown in (c). The rods consist of Si or GaAs (full black circles)
or polymer (open red circles).
Example A represents a close to optimal structure with
{\em inter-site} location of the $\alpha$-resonator whose
resonance frequency lies close to the edge $k=\pm \pi/s$
of the passing band; example B represents a sub-optimal
structure with an {\em inter-site} location of the $\alpha$-resonator
whose resonance frequency lies near the center of the
passing band; example C represents a sub-optimal but commonly used
structure with an {\em on-site} location of the
$\alpha$-resonator whose resonance frequency lies near the
center of the passing band. Closed circles in (a) indicate
frequencies with $T(\omega)=80\%$ that are used for achieving
high-contrast bistability in (b). Red circles in (c) indicate
positions of the nonlinear polymer rods with  $\varepsilon_{\alpha}=2.56$.
Other parameters of the 2D photonic crystal are described in
Appendix~\ref{sec:example-PhC}.} \label{fig:example-polymer-wg}
\end{figure}

First, we consider a waveguide created by removing {\em every second
rod} ($s=2a$) in a straight line of rods coupled to a nonlinear
resonator $\alpha$ created by replacing a single rod of the
two-dimensional lattice with a highly-nonlinear polymer rod.
The corresponding structure is schematically shown in the insets
in Fig.~\ref{fig:example-air-wg}.
The resonant frequency of the polymer-rod resonator lies very
close to the edge $k=\pm \pi/s$ of the waveguide passing band,
and can be tuned by changing the linear dielectric constant
$\varepsilon_{\alpha}$ of the rod.

In Fig.~\ref{fig:example-air-wg}(a) and (b), respectively, we
display the transmission spectra for both on-site and inter-site
positions of the side-coupled resonator for three different values
of resonator dielectric constant $\varepsilon_{\alpha}$. We notice
that in the case of the on-site position of the resonator the
transmission coefficient $T(\omega)$ remains below the critical
value of $T=75\%$ required for bistable switching operation for
all frequencies $\omega$ below the resonance frequency
$\omega_{\text{res}}$. The condition $\omega<\omega_{\text{res}}$
corresponds to the condition $(\sigma(\omega) \cdot \Jalpha)>0$
which should be satisfied to realize nonmonotonic dependencies of
the nonlinear transmission shown in Fig.~\ref{fig:T-L1A1-sigma}).
{\em Therefore, this on-site system cannot exhibit bistability}.

On the other hand, bistability may be realized for the inter-site
position of the side-coupled resonator for which, in a full
agreement with our analysis presented above, the transmission
remains perfect at the band edge $k=\pm \pi/s$ and the quality
factor $Q$ increases as the resonant frequency approaches this
band edge. In Fig.~\ref{fig:example-polymer-wg}(b) (example A) we
show that in this case bistable transmission indeed occurs for the
frequency marked by a filled circle in
Fig.~\ref{fig:example-polymer-wg}(a). This  corresponds to
$T(\omega)=80\%$, i.e., the choice $\sigma^2(\omega)=4$ for the
detuning parameter.

We want to emphasize that the large value of the quality factor
(\ref{Q-factor-inter}) for the inter-site structure at $k(\omega)$
close to $\pm \pi/s$ leads to very low bistability thresholds as
compared to the cases of on-site coupled and continous waveguide
coupled structures. This is illustrated in examples B and C of
Fig.~\ref{fig:example-polymer-wg}: Relative to the waveguide
design in example A, the design in example B moves the resonance
frequency deeper into the passing band thus decreasing the quality
factor (\ref{Q-factor-inter}). Nevertheless, the inter-site
coupled example B still exhibits a much smaller bistability
threshold than the on-site coupled system with the same waveguide
design in example C. This is caused by (usually) much smaller
waveguide-resonator coupling and, accordingly, much larger $Q$ in
the inter-site structures as compared to the on-site structures.

Summarizing, the inter-site structure of the resonant
waveguide-resonator interaction schematically shown in
Fig.~\ref{fig:struct}(c) allows to achieve much higher values for
the linear quality factor $Q$. As a consequence, much smaller
bistability threshold intensities for the nonlinear transmission
are obtained. To employ these advantages, the wavevector
$k(\omega_{\text{res}})$ of the guided mode at the resonance
frequency $\omega_{\text{res}}$,
Eq.~(\ref{sigma-local-linear-inter}), should be as close as
possible to $\pi/s$. This requirement coincides with the condition
of a very small group velocity in the waveguide and, in contrast
to the continuous-waveguide  and on-site structures depicted in
Figs.~\ref{fig:struct}(a,b), provides us with a possibility to
create low-threshold all-optical switching devices based on
slow-light photonic crystal waveguides.

\section{Conclusions}

We have presented a detailed analysis of PhC 
waveguides side-coupled to Kerr nonlinear resonators which may
serve as a basic element of active photonic-crystal circuitry.
First, we have extended the familiar approach based on standard
coupled-mode theory and derived explicit analytical expressions
for the bistability thresholds and transmission coefficients
related to light switching in such structures. Our results reveal
that, from the point of view of bistability contrast (a small
difference between two threshold intensities and robustness of
switching) the best conditions for bistability are realized for
those parameter values for which the dimensionless detuning
parameter $\sigma(\omega)$ is close to $\sqrt{5}$. Practically, this
corresponds to the choice of operation frequencies for which the
linear light transmission is close to 83\%.

We have pointed out that the conventional coupled-mode theory does
not allow to describe the light transmission near the band edges,
and we have developed an improved semi-analytical approach based
on the effective discrete equations derived in the framework of a
consistent Green's function formalism. This approach is ideally
suited for a qualitative and semi-quantitative description of
photonic-crystal devices that involve a discrete set of
small-volume cavities. We have shown that this novel approach
allows to adequately describe light transmission in the
waveguide-resonator structures near the band edges. Specifically,
we have demonstrated that while the transmission coefficient
vanishes at both spectral edges for the on-site coupled structure
(see Fig.~\ref{fig:struct}(b)), light transmission remains perfect
at one band edge for the inter-site coupled structure (see
Fig.~\ref{fig:struct}(c)). These features allow a significant
enhancement of the resonator quality factor and, accordingly, a
substantial reduction of the bistability threshold. As a
consequence, we refer to this type of nonlinearity enhancement as
a {\em geometric enhancement}. The possibility of such enhancement
is a direct consequence of the discreteness of the photonic
crystal waveguide and is in a sharp contrast to similar resonant
systems based on ridge waveguides. The potential of this novel
type of the nonlinearity enhancement may be regarded as an
additional argument to support the application of photonic-crystal
devices in integrated photonic circuits.

In addition, we would like to emphasize that the engineering of
the geometry of photonic-crystal based devices such as that
presented in Fig.~\ref{fig:struct}(c) becomes extremely useful for
developing novel concepts of all-optical switching in the {\em
slow-light regime} of PhC waveguides which may have
much wider applications in nanophotonics and is currently under
active experimental research \cite{Vlasov-slow-light}.

We believe that the basic concept of the geometric enhancement of
nonlinear effects based on the discrete nature of photonic-crystal
waveguides will be useful in the study of more complicated devices
and circuits and, in particular, for various slow-light
applications. For instance, this concept may be applied to the
transmission of a side-coupled resonator placed between two
partially reflecting elements embedded into the photonic-crystal
waveguide where sharp and asymmetric line shapes have been
predicted with associated variations of the transmission from 0\%
to 100\%  over narrow frequency ranges~\cite{Fan:2002-908:APL}.
Similarly, the concept can be extended to a system of cascaded
cavities~\cite{Lin:2005-165330:PRB} and three-port channel-drop
filters~\cite{Kim:2004-5518:OE}, optical delay
lines~\cite{Wang:2003-66616:PRE}, systems of two nonlinear
resonators with a very low bistability
threshold~\cite{Maes:2005-1778:JOSB}, etc.

\acknowledgments

S.F.M. and K.B. acknowledge a support from the Center for
Functional Nanostructures of the Deutsche Forschungsgemeinschaft
within the project A1.1. S.F.M. also acknowledges a support from
the Organizers of the PECS-VI Symposium
(http://cmp.ameslab.gov/PECSVI/), where some of these results have
been presented for the first time. The work of Y.K. and A.E.M. was
supported by the Australian Research Council through the Center of
Excellence Program.

\appendix
\section{Calculation of the model parameters and examples}

\subsection{Coupling coefficients for two-dimensional photonic crystals}
\label{sec:example-PhC}

To obtain deeper insight into the basic properties of the
effective discrete equations (\ref{discrete-eqs}), we should know
how the coupling coefficients $D_n(\omega)$, $\kappa_n(\omega)$,
and $V_{n,m}(\omega)$ depend on frequency $\omega$. As an
illustration, we consider a two-dimensional model of a photonic
crystal consisting of a square lattice (lattice spacing $a$) of
infinitely long dielectric rods (see Refs.~
\cite{Fan:1998-960:PRL,Soljacic:2002-55601:PRE,Yanik:2003-2739:APL}
and also references [7-16] in
Ref.~\cite{Mingaleev:2002-2241:JOSB}). We study light propagation
in the plane of periodicity, assuming that the rods have a radius
$r=0.18a$ and a dielectric constant of
$\varepsilon_{\text{rod}}=11.56$ (GaAs or Si at the
telecommunication wavelength $\lambda \sim 1.55$ $\mu$m). For
light with the electric field polarized along the rods
($E$-polarized light), this photonic crystal exhibits a large
(38\% of the center frequency) photonic bandgap that extends from
$\omega = 0.303\, (2\pi c/a)$ to $\omega = 0.444\, (2\pi c/a)$

\begin{figure}[tbp]
\centerline{\scalebox{.37}{\includegraphics[clip]{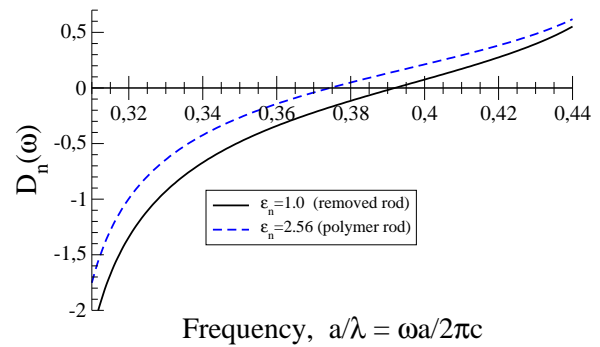}}}
\caption{(Color online) Frquency dependence of the detuning
coefficient $D_n(\omega)$ for the 2D photonic crystal described in
Appendix~\ref{sec:example-PhC}, for two types of resonators:
Removing a single rod ($\varepsilon_n=1.0$; solid line) leads to a
localized mode at $\omega_n=0.392$, while replacing a single rod
by a geometrically identical rod made of polymer
($\varepsilon_n=2.56$, dashed line) leads to a localized mode at
$\omega_n=0.374$.} \label{fig:Dn}
\end{figure}

\begin{figure}[tbp]
\centerline{\scalebox{.37}{\includegraphics[clip]{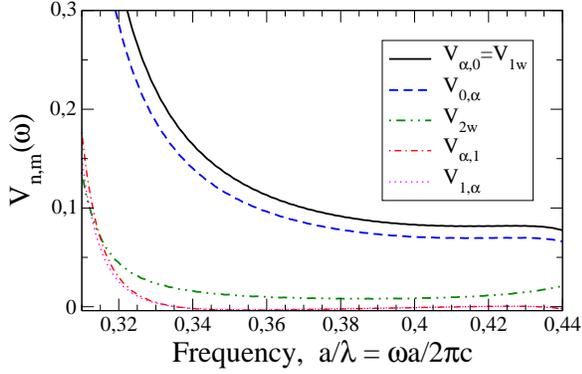}}}
\caption{(Color online)  Frequency dependence of the coupling
coefficients $V_{n,m}(\omega)$ for the 2D photonic crystal described
in Appendix~\ref{sec:example-PhC} with a on-site side-coupled
waveguide-resonator system shown in Fig.~\ref{fig:struct}-b. The
notations are the same as those in Eq.~(\ref{discrete-eqs-Fano})
and we assume that the waveguide is created by removing every
second rod in a row (located at $\vec{R}_n=2\vec{a}_1 n$) whereas
the on-site resonator is created by replacing a single rod at
$\vec{R}_{\alpha}=-2\vec{a}_2$ with a polymer rod
($\varepsilon_n=2.56$) of the same radius. The dispersion
relation for such a waveguide is displayed in Fig.~\ref{fig:disp-wg-crow1}.}
\label{fig:Vnm}
\end{figure}

Our task is to evaluate the coupling coefficients $D_n(\omega)$,
$\kappa_n(\omega)$, and $V_{n,m}(\omega)$ using
Eqs.~(\ref{V-coeffs})--(\ref{K-coeffs}) with the Green's function
$\Green$ calculated by the method described earlier in
Refs.~\cite{Mingaleev:2000-5777:PRE,Mingaleev:2001-5474:PRL}. The
results of these calculations are displayed in
Figs.~\ref{fig:Dn}--\ref{fig:Vnm}.

\subsection{Isolated optical resonators}

For the case of an isolated ($V_{n,m}=0$) linear
($\chi^{(3)}_{n}=0$) optical resonator at the site $n$,
Eq.~(\ref{discrete-eqs}) takes a simplest possible form,
$D_{n}(\omega) A_n = 0$. In this case, we only need to know the
dimensionless frequency detuning coefficient, $D_n(\omega)$. In
Fig.~\ref{fig:Dn} we plot $D_n(\omega)$ for two types of resonators: a
resonator created by removing a single rod and a resonator created
by replacing a single rod with a polymer rod of the same radius
and $\varepsilon_{n}=2.56$. Introducing the dimensionless
frequency $\tilde{\omega}=a/\lambda \equiv (\omega a/2\pi c)$, we
can express these coefficients, with a very good accuracy in the
range $0.36 \leq \tilde{\omega} \leq 0.41$, by the following cubic
dependencies: $D_n(\omega) = 9.426 (\tilde{\omega} -
\tilde{\omega}_{\text{n}}) -10.889(\tilde{\omega} -
\tilde{\omega}_{\text{n}})^2 + 840.36(\tilde{\omega} -
\tilde{\omega}_{\text{n}})^3$ with
$\tilde{\omega}_{\text{n}}=0.3919$, for the removed rod, and
$D_n(\omega) = 9.047 (\tilde{\omega} - \tilde{\omega}_{\text{n}})
-49.555(\tilde{\omega} - \tilde{\omega}_{\text{n}})^2 +
770.14(\tilde{\omega} - \tilde{\omega}_{\text{n}})^3$ with
$\tilde{\omega}_{\text{n}}=0.3744$, for the replaced rod.

The resonator mode can only be excited at the resonator frequency
$\omega_n$, which is determined by the equation $D_{n}(\omega_n)=0$.
Fig.~\ref{fig:Dn} suggests that  changing the dielectric constant
of the resonator $\varepsilon_n$ allows to tune the frequency
$\omega_n$. In all cases, in the vicinity of the resonator frequency
$\omega_n$, the coupling coefficient $D_{n}(\omega)$ can be approximately
expanded into the Taylor series with a linear dependence
\begin{eqnarray}\label{Taylor-series-for-D}
D_{n}(\omega) \simeq \frac{\omega-\omega_{n}} {\nu_n \,
\omega_{n}} \; , \quad \nu_n = \frac{1}{\omega_{n}
D_{n}'(\omega_{n})} \; ,
\end{eqnarray}
where we have introduced a dimensionless parameter $\nu_n$ which
describes the resonator sensitivity to a change of the dielectric
constant. For our example of a polymer-rod resonator, we find
$\nu_n \approx 0.295$.

\begin{figure}[tbp]
\centerline{\scalebox{.37}{\includegraphics[clip]{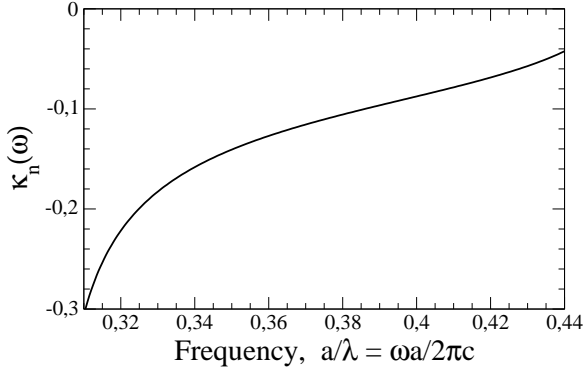}}}
\caption{Frequency dependence of the nonlinear feedback parameter
$\kappa_n(\omega)$ for the 2D photonic crystal described in
Appendix~\ref{sec:example-PhC}. The nonlinear resonator is created
by replacing a single rod with a polymer rod of the same radius
which supports at $\varepsilon_n=2.56$ a localized mode with
frequency $\omega_n=0.374$.} \label{fig:Kn}
\end{figure}

When the $n$-th resonator is nonlinear (i.e., $\chi^{(3)}_{n} \neq
0$), Eq.~(\ref{discrete-eqs}) reduces to the equation
$D_{n}(\omega) A_n = \kappa_{n}(\omega) \chi^{(3)}_n |A_n|^2 A_n$
with the new important coefficient --- {\em nonlinear feedback
parameter} $\kappa_n(\omega)$. In Fig.~\ref{fig:Kn}, we depict the
frequency dependence of $\kappa_n(\omega)$ for the case of a
nonlinear polymer resonator. In the frequency range $0.36 \leq
\tilde{\omega} \leq 0.41$, this behavior can be approximated as
$\kappa_n(\omega) = -0.111 + 1.005 (\tilde{\omega} -
\tilde{\omega}_{\text{n}}) -5.501(\tilde{\omega} -
\tilde{\omega}_{\text{n}})^2 + 85.57 (\tilde{\omega} -
\tilde{\omega}_{\text{n}})^3$ with
$\tilde{\omega}_{\text{n}}=0.3744$. Therefore, in the vicinity of
the resonator's frequency, $\omega_{n}$, we may assume that
$\kappa_{n}(\omega) \approx -0.111$ is constant and can rewrite
Eq.~(\ref{discrete-eqs}) according to
\begin{eqnarray}
\label{discrete-eqs-single:sol} |A_n|^2 =
\frac{D_{n}(\omega)}{\kappa_{n}(\omega) \chi^{(3)}_n} \approx
\frac{D_{n}'(\omega_n)}{\kappa_{n}(\omega_n) \chi^{(3)}_n}
(\omega-\omega_n).
\end{eqnarray}
The solution of the above equation gives us the dependence of the
resonator frequency $\omega_{\text{res}}$ on the resonator's mode
intensity $|A_n|^2$ as
\begin{eqnarray}\label{discrete-eqs-single:sol-omega}
\omega_{\text{res}} \approx \omega_n \left(1 + \kappa_n \nu_{n} \,
\chi^{(3)}_n \, |A_n|^2 \right) \; .
\end{eqnarray}
Here, we have used the notation $\kappa_n=\kappa_{n}(\omega_n)$.
As we see, the nonlinear sensitivity of the resonator at the site
$n$ is a product of its nonlinear feedback parameter, $\kappa_n$,
the sensitivity to a change of the dielectric constant, $\nu_n$,
and the Kerr susceptibility of  material, $\chi^{(3)}_n$. The sign
of this product defines the direction of the resonator frequency
shift. In particular, for the polymer resonator used in
Figs.~\ref{fig:Dn}--\ref{fig:Kn}, we obtain a rather small shift,
$\kappa_n \nu_n \approx -0.033$ which indicates that for
$\chi^{(3)}_n>0$ the resonator frequency decreases as the light
intensity grows. Designing optical resonators with larger
$\kappa_{n}$ or $\nu_n$, may allow to enhance their nonlinear
properties for a given material with Kerr nonlinearity
$\chi^{(3)}_n$.

\subsection{Straight waveguides}

Now let us consider an array of identical coupled cavities
separated by the distance $s=|\vec{s}|$ which create a straight
photonic-crystal waveguide depicted in
Figs.~\ref{fig:struct}(b,c). Before proceeding, we would like to
emphasize that our analysis can equally well be applied to the
coupled-resonator optical waveguides (CROWs) suggested in
Ref.~\cite{Yariv:1999-711:OL}. If we neglect nonlinear effects
(assuming that either the waveguide cavities are linear,
$\chi^{(3)}_n=0$, or the light intensity in the waveguide remains
sufficiently small), Eq.~(\ref{discrete-eqs}) reduces to
\begin{eqnarray}\label{discrete-eqs-wg}
D_{\text{w}}(\omega) A_n =
\sum_{j=1}^{\infty} V_{j\text{w}}(\omega) (A_{n+j}+A_{n-j}) \; .
\end{eqnarray}
Here we have defined, similar to Eq.~(\ref{discrete-eqs-Fano}),
$D_{\text{w}}(\omega) \equiv D_{n}(\omega)$ and
$V_{j\text{w}}(\omega) \equiv V_{n,n \pm j}(\omega)$ which are
identical for all $n$.

In Fig.~\ref{fig:Vnm} we plot the frequency dependencies of
$V_{1\text{w}}(\omega)$ and $V_{2\text{w}}(\omega)$ for a
photonic-crystal waveguide created by removing every second rod in
a row, either with $\vec{s}=2\vec{a}_1$ or with
$\vec{s}=2\vec{a}_2$. In the vicinity of the polymer-rod resonator
frequency, the coupling coefficients are to lowest order constant:
$V_{1\text{w}} \approx 0.096$ and $V_{2\text{w}} \approx 0.0086$ .
In the general case, our calculations show that the coefficients
$V_{j\text{w}}(\omega)$ decay nearly exponentially with $j$. In
terms of frequency, they take on a constant value at the central
passing band frequency and grow rapidly towards the low-frequency
bandgap edge.

\begin{figure}[tbp]
\centerline{\scalebox{.37}{\includegraphics[clip]{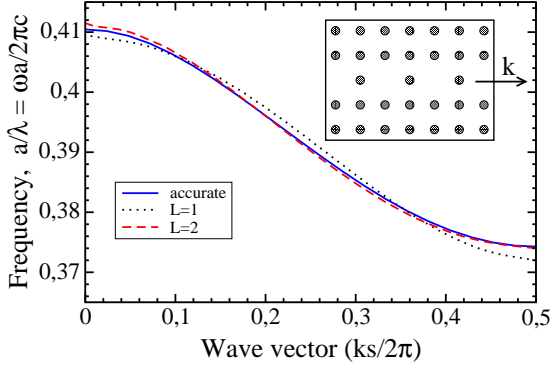}}}
\caption{(Color online) Dispersion relation for a photonic crystal
waveguide created by removing every second rod in a row
($\vec{s}=2\vec{a}_1$) in the 2D photonic crystal described in
Appendix~\ref{sec:example-PhC}. Numerical exact results (solid
line) are calculated with the super-cell plane-wave method
\cite{Johnson:2001-173:OE} and the approximate results are
obtained from Eq.~(\ref{k-wave-nonloc}) with $L=1$ (dotted line)
and $L=2$ (dashed line), using the coupling coefficients from
Fig.~\ref{fig:Vnm}.}
\label{fig:disp-wg-crow1}
\end{figure}

According to the Floquet-Bloch theorem,
Eq.~(\ref{discrete-eqs-wg}) has a solution $A_n=A_0 \exp[\pm
ik(\omega)sn]$ with an arbitrary complex amplitude $A_0$. The
corresponding dispersion $k(\omega)$ is determined by the equation
\begin{eqnarray}\label{k-wave-nonloc}
D_{\text{w}}(\omega) =
\sum_{j=1}^{L} 2 V_{j\text{w}}(\omega) \cos
\left[k(\omega) s \, j \right] \; ,
\end{eqnarray}
where we assume that the coupling coefficients
$V_{j\text{w}}(\omega)$ vanish for all $j$ above $L$. As a matter
of fact, our studies indicate that sufficiently accurate results
can be obtained already for $L \sim 4a/s$. In
Fig.~\ref{fig:disp-wg-crow1} we plot the dispersion relation for a
2D model photonic-crystal waveguide and compare it with exact
numerical results calculated by the super-cell plane-wave
method~\cite{Johnson:2001-173:OE}. For this case, even the
simplest tight-binding approximation (i.e., at $L=1$) gives quite
satisfactory results.

In the tight-binding approximation ($L=1$) the dispersion relation
can be described by the following simple expression
\begin{eqnarray}\label{k-wave-TB}
\cos \left[ k(\omega) s \right]
= \frac{D_{\text{w}}(\omega)}{2 V_{1\text{w}}(\omega)}
\simeq \frac{\omega-\omega_{\text{w}}}{\omega_{\text{w}}\Delta_{\text{w}}} \; ,
\end{eqnarray}
where $\omega_{\text{w}}$ is the resonance frequency of the
waveguide cavities. Furthermore, we have the dimensionless
parameter
\begin{eqnarray}\label{Delta-A}
\Delta_{\text{w}}=\frac{2 V_{1\text{w}}(\omega_{\text{w}})}
{\omega_{\text{w}} D'_{\text{w}}(\omega_{\text{w}})}
= 2 V_{1\text{w}} \nu_{\text{w}} \; ,
\end{eqnarray}
with $V_{1\text{w}} \equiv V_{1\text{w}}(\omega_{\text{w}})$ and
$\nu_{\text{w}}$ defined by Eq.~(\ref{Taylor-series-for-D}), that
equals half-bandwidth of the waveguide transmission band. This
band extends from $\omega_{\text{w}}(1-\Delta_{\text{w}})$ to
$\omega_{\text{w}}(1+\Delta_{\text{w}})$. For our example of
photonic crystal waveguide, we find $\Delta_{\text{w}} \approx
0.052$, i.e., its bandwidth is about $10\%$.

\section{Effect of long-range interactions}

\subsection{Effects of nonlocal dispersion}

As follows from the results of Sec.~\ref{sec:on-site} above, the
local-coupling approximation provides us with an excellent
qualitative analysis of the structure shown in
Fig.~\ref{fig:struct}(b). However, certain physically important
effects may be missed in this approximation. A detailed analysis
of the effects of nonlocal coupling was performed in
Ref.~\cite{Miroshnichenko:2005-36626:PRE}, so that here we may
discuss this issue very briefly, and may specify it directly to
photonic-crystal devices.

In Fig.~\ref{fig:transm-crow1-fano2}, we provide a comparison of
$T(\omega)$ calculated from Eq.~(\ref{sigma-local-linear}) in the
local-coupling approximation with the exact numerical results for
the structure shown in Fig.~\ref{fig:struct}(b) for the model
photonic crystal described in Appendix~\ref{sec:example-PhC}. The
results suggest that the local-coupling approximation introduces a
frequency shift for the band edges which agrees well with the
corresponding frequency shift in the dispersion relation shown in
Fig.~\ref{fig:disp-wg-crow1}.

In addition, the resonance frequency is also shifted; it is not
equal to $\omega_{\alpha}$ but is slightly larger. In principle,
this shift can be produced by two effects: (i) a long-range
coupling between cavities inside the waveguide and (ii) a
long-range coupling between the waveguide and the side-coupled
resonator. First, we explore the former possibility.

Solving Eqs.~(\ref{discrete-eqs-Fano})--(\ref{En-asymptotic}) for
$L=2$, we obtain the transmission and reflection coefficients
(\ref{tr-Fano})--(\ref{tr-Fano-abs}) with the detuning parameter
\begin{widetext}
\begin{eqnarray}\label{sigma-linear-L2}
\sigma(\omega) = 2 \sin (ks) \, \cdot \frac{D_{\alpha} \left[
V_{1\text{w}}^3 + 3 D_{\text{w}} V_{1\text{w}} V_{2\text{w}} + 3
V_{1\text{w}} V_{2\text{w}}^2 + 2 V_{2\text{w}}^3 \cos(3ks) \right]
- V_{0,\alpha} V_{\alpha,0} V_{1\text{w}} V_{2\text{w}} }
{V_{0,\alpha} V_{\alpha,0} \left( V_{1\text{w}}^2 - V_{2\text{w}}^2
+ D_{\text{w}} V_{2\text{w}} \right)} \; ,
\end{eqnarray}
\end{widetext}
where all the coefficients are assumed to be frequency-dependent
analogous to Eqs.~(\ref{mu})-(\ref{sigma-local-linear}). The
waveguide dispersion $k(\omega)$ is now calculated from
Eq.~(\ref{k-wave-nonloc}) with $L=2$.

Fig.~\ref{fig:transm-crow1-fano2} shows that the transmission
calculated from Eq.~(\ref{sigma-linear-L2}) is much closer to
the exact numerical results. In fact, the nominator of
Eq.~(\ref{sigma-linear-L2}) indicates that, indeed, the resonance
frequency  is slightly shifted from the value $\omega_{\alpha}$,
and that this shift is proportional to $V_{2\text{w}}$. Since
$V_{2\text{w}}$ is always much smaller than $V_{1\text{w}}$
(see Fig.~\ref{fig:Vnm}), we can safely neglect all the terms
proportional to $V_{2\text{w}}^n$ with $n \geq 2$, and obtain
the resonance frequency according to
\begin{eqnarray}\label{omega-res-L2}
\omega_{\text{res}} \approx \omega_{\alpha} \left(1 +
\frac{ V_{0,\alpha} V_{\alpha,0} V_{1\text{w}} V_{2\text{w}} }
{(V_{1\text{w}}^3 + 3 D_{\text{w}} V_{1\text{w}}
V_{2\text{w}})} \, \nu_{\alpha} \right) \; .
\end{eqnarray}
Here, the values of all coefficients are calculated at the
frequency $\omega_{\alpha}$, and $\nu_{\alpha}$ is defined by
Eq.~(\ref{Taylor-series-for-D}).

In addition to the shift of the resonance frequency, a perfect
transmission may occur at the frequencies for which the
denominator in Eq.~(\ref{sigma-linear-L2}) vanishes:
\begin{eqnarray}\label{omega-perfect-L2}
V_{2\text{w}} = V_{1\text{w}} \left( \frac{\cos (ks) \pm
\sqrt{1+\cos^2 (ks) -2 \cos(2ks)}}{1-2\cos(2ks)} \right) \; .
\end{eqnarray}
However, an analysis reveal that Eq.~(\ref{omega-perfect-L2})
has solutions only when $|V_{2\text{w}}(\omega)|$ exceeds
$|V_{1\text{w}}(\omega)|/2$, a condition that appears to be
impossible to realize in realistic photonic crystals.

\subsection{Effects of nonlocal coupling}

Another possible reason for a shift of the resonance frequency is
a nonlocal coupling between the waveguide cavities and the
side-coupled resonator $\alpha$. Here, we discuss this effect in
the framework of the tight-binding approximation for the waveguide
dispersion (i.e., $L=1$) to distinguish it from the other type of
nonlocal effects discussed in the previous sub-section. We assume
that $V_{j,\alpha}(\omega)=V_{\alpha,j}(\omega) = 0$ for all $j
\geq 2$, and take into account that, for the symmetric structure
shown in Fig.~\ref{fig:struct}(b), the coupling coefficients are:
$V_{-1,\alpha}(\omega) \equiv V_{1,\alpha}(\omega)$ and
$V_{\alpha,-1}(\omega) \equiv V_{\alpha,1}(\omega)$. Then, we
obtain a solution of
Eqs.~(\ref{discrete-eqs-Fano})--(\ref{En-asymptotic}) in the form
of Eqs.~(\ref{tr-Fano})--(\ref{tr-Fano-abs}) with the detuning
parameter defined as
\begin{widetext}
\begin{eqnarray}\label{sigma-linear-L1-A1}
\sigma(\omega) = 2 \sin [ks] \, \cdot \frac{D_{\alpha} V_{1\text{w}}
+ V_{0,\alpha} V_{\alpha,1} + V_{\alpha,0} V_{1,\alpha} + 2
V_{\alpha,1} V_{1,\alpha} \cos (ks)} {\bigl[ V_{\alpha,0} + 2
V_{\alpha,1} \cos (ks) \bigr] \bigl[ V_{0,\alpha} + 2 V_{1,\alpha}
\cos (ks) \bigr]} \; .
\end{eqnarray}
\end{widetext}
Here, all coefficients are assumed to be frequency-dependent,
similar to Eqs.~(\ref{mu})-(\ref{sigma-local-linear}).
Furthermore, the waveguide dispersion $k(\omega)$ is calculated
again from Eq.~(\ref{k-wave-TB}).

Eq.~(\ref{sigma-linear-L1-A1}) suggests that in this case the
resonance frequency becomes slightly shifted from the value
$\omega_{\alpha}$, and this shift is proportional to the values of
$V_{1,\alpha}$ and $V_{\alpha,1}$, which for our example (see
Fig.~\ref{fig:transm-crow1-fano2}) are equal to
$V_{\alpha,1}=-0.0026$ and $V_{1,\alpha}=-0.0022$. Assuming that
these coupling coefficients are always much smaller than
$V_{0,\alpha}$ and $V_{\alpha,0}$ (cf. $V_{\alpha,0}=0.096$ and
$V_{0,\alpha}=0.082$), we obtain for the resonance frequency
\begin{eqnarray}\label{omega-res-L1-A1}
\omega_{\text{res}} \approx \omega_{\alpha} \left( 1 -
\frac{ V_{0,\alpha} V_{\alpha,1} + V_{\alpha,0} V_{1,\alpha}}
{V_{1\text{w}}} \, \nu_{\alpha} \right)\; .
\end{eqnarray}
Here, the coefficients are calculated at the resonance frequency
$\omega_{\alpha}$, and $\nu_{\alpha}$ is defined by
Eq.~(\ref{Taylor-series-for-D}). For the example shown in
Fig.~\ref{fig:transm-crow1-fano2}, this frequency shift is much
smaller than that described by Eq.~(\ref{omega-res-L2}) because in
this case the values of $V_{\alpha,1}$ and $V_{1,\alpha}$ are 3.3
times smaller than the value of $V_{2\text{w}}$.

Due to this long-range coupling, there appears a possibility of
perfect light transmission, as discussed in
Ref.~\cite{Miroshnichenko:2005-36626:PRE}, but only in the case when
$|V_{\alpha,1}(\omega)|$ exceeds $|V_{\alpha,0}(\omega)|/2$ or
$|V_{1,\alpha}(\omega)|$ exceeds $|V_{0,\alpha}(\omega)|/2$.
Again, such a scenario appears to be impossible to realize in
realistic photonic crystal structures.



\begin{thebibliography}{}

\bibitem{Gibbs:1985:Book}
H.M. Gibbs, {\em Optical bistability: Controlling light with light}
(Academic Press, Orlando, 1985).

\bibitem{PhCs}
K. Busch, S. L{\"o}lkes, R.B. Wehrspohn, and H. F{\"o}ll (Eds.),
{\em Photonic Crystals: Advances in Design, Fabrication, and
Characterization} (Wiley-VCH, Berlin, 2004); K.~Inoue and K.~Ohtaka
(Eds.), {\em Photonic Crystals: Physics, Fabrication and
Applications} (Springer, Berlin, 2004).

\bibitem{Noda:2000-608:NAT}
S. Noda, A. Chutinan, and M. Imada, Nature {\bf 407},  608  (2000).

\bibitem{Chutinan:2001-2690:APL}
A. Chutinan, M. Mochizuki, M. Imada, and S. Noda, Appl. Phys. Lett.
{\bf 79},  2690  (2001).

\bibitem{Asano:2003-407:APL}
T. Asano, B.~S. Song, Y. Tanaka, and S. Noda, Appl. Phys. Lett. {\bf
83}, 407  (2003).

\bibitem{Imada:2002-873:JLT}
M. Imada, S. Noda, A. Chutinan, M. Mochizuki, and T. Tanaka, J.
Lightwave Techn. {\bf 20}, 873 (2002)

\bibitem{Smith:2001-1487:APL}
C.J.M. Smith {\it et~al.}, Appl. Phys. Lett. {\bf 78},  1487 (2001).

\bibitem{Seassal:2002-811:IQE}
C. Seassal {\it et~al.}, IEEE J. Quantum Electron. {\bf 38},  811
(2002).

\bibitem{Notomi:2005-2678:OE}
M. Notomi {\it et~al.}, Opt. Express {\bf 13},  2678  (2005).

\bibitem{Barclay:2005-801:OE}
P.E. Barclay, K. Srinivasan, and O. Painter, Opt. Express {\bf 13},
801  (2005).

\bibitem{haus}
H.A. Haus, {\em Waves and Fields in Optoelectronics},
(Prentice–Hall, Englewood Cliffs, NJ, 1984).

\bibitem{Haus:1992-205:IQE}
H.~A. Haus and Y. Lai, IEEE J. Quantum Electron. {\bf 28}, 205
(1992).

\bibitem{Fano:1961-1866:PREV}
U. Fano, Phys. Rev. {\bf 124}, 1866 (1961).

\bibitem{Anderson:1961-41:PREV}
P.~W. Anderson, Phys. Rev. {\bf 124}, 41 (1961).

\bibitem{Fan:1998-960:PRL}
S.~H. Fan, P.~R. Villeneuve, and J.~D. Joannopoulos, Phys. Rev.
Lett.
  {\bf 80},  960  (1998).

\bibitem{Xu:2000-7389:PRE}
Y. Xu, Y. Li, R.~K. Lee, and A. Yariv, Phys. Rev. E {\bf 62}, 7389
(2000).

\bibitem{Soljacic:2002-55601:PRE}
M. Soljacic, M. Ibanescu, S. G. Johnson, Y. Fink, and J. D.
Joannopoulos, Phys. Rev. E {\bf 66}, 055601(R) (2002).

\bibitem{Yanik:2003-2739:APL}
M.~F. Yanik, S.~H. Fan, and M. Soljacic, Appl. Phys. Lett. {\bf
83},  2739  (2003).

\bibitem{Chak:2006-035105:PRE}
P. Chak, S. Pereira, and J.~E. Sipe, Phys. Rev. B {\bf 73}, 035105
(2006).

\bibitem{Cowan:2003-46606:PRE}
A.~R. Cowan and J.~F. Young, Phys. Rev. E {\bf 68},  046606
(2003).

\bibitem{Cowan:2005-R41:SST}
A.~R. Cowan and J.~F. Young, Semicond. Sci. Technol. {\bf 20},
R41  (2005).

\bibitem{Miroshnichenko:2005-36626:PRE}
A.~E. Miroshnichenko, S.~F. Mingaleev, S. Flach, and Yu.~S. Kivshar, Phys. Rev.
  E {\bf 71},  036626  (2005).

\bibitem{Miroshnichenko:2005-56611:PRE}
A.~E. Miroshnichenko and Yu.~S. Kivshar,
Phys. Rev. E {\bf 72},  056611  (2005).

\bibitem{Miroshnichenko:2005-3969:OE}
A.~E. Miroshnichenko and Yu.~S. Kivshar, Opt. Express \textbf{13},
3969 (2005).

\bibitem{Mingaleev:2002-2241:JOSB} S.F. Mingaleev and Yu.S. Kivshar,
J. Opt. Soc. Am. B \textbf{19}, 2241 (2002).

\bibitem{Mingaleev:2002-231:OL}
S.~F. Mingaleev and Yu.~S. Kivshar, Opt. Lett. {\bf 27},  231  (2002).

\bibitem{McGurn:1999:PLA}
A.~R. McGurn, Phys. Lett. A {\bf 251}, 322 (1999);
Phys. Lett. A {\bf 260}, 314 (1999).

\bibitem{Mingaleev:2000-5777:PRE}
S.~F. Mingaleev, Yu.~S. Kivshar, and R.~A. Sammut, Phys. Rev. E {\bf 62},  5777
  (2000).

\bibitem{Mingaleev:2001-5474:PRL}
S.~F. Mingaleev and Yu.~S. Kivshar, Phys. Rev. Lett. {\bf 86},  5474  (2001).

\bibitem{Waks:2005-5064:OE}
E. Waks and J. Vuckovic, Opt. Express {\bf 13},  5064  (2005).

\bibitem{Sheik-Bahae:1991-1296:IQE}
M. Sheik-Bahae, D.~C. Hutchings, D.~J. Hagan, and E.~W. Van Stryland,
IEEE J. Quantum Electron. {\bf 27}, 1296 (1991).

\bibitem{Samoc:1995-1241:OL}
A. Samoc, M. Samoc, M. Woodruff, and B. Luther-Davies,
Opt. Lett. {\bf 20}, 1241 (1995).

\bibitem{Mingaleev:2004-2858:OL}
S. F. Mingaleev, M. Schillinger, D. Hermann, and K. Busch,
Opt. Lett. {\bf 29}, 2858 (2004).

\bibitem{Schillinger:2005-324:SPIE}
M. Schillinger, S. F. Mingaleev, D. Hermann, and K. Busch,
Proc. of SPIE {\bf 5733} -- Photonic Crystal Materials and Devices III,
324 (2005).

\bibitem{Jiao:2005-1875:IPTL}
Y. Jiao, S. F. Mingaleev, M. Schillinger, D. A. B. Miller,
S. Fan, and K. Busch, IEEE Photonics Technol. Lett.
{\bf 17}, 1875 (2005).

\bibitem{Heijden:2006-161112:APL}
R. van der Heijden et al., Appl. Phys. Lett. \textbf{88}, 161112
(2006).

\bibitem{Ferrini:2006-1238:OL}
R. Ferrini et al., Opt. Lett. \textbf{31}, 1238 (2006).

\bibitem{Akahane:2003-944:NAT}
Y. Akahane, T. Asano, B.-S. Song, and S. Noda, Nature {\bf 425},
944 (2003).

\bibitem{Busch:2003-R1233:JPCM}
K. Busch, S. F. Mingaleev, A. Garcia-Martin, M. Schillinger, and
D. Hermann,
J. Phys.: Condens. Matter. \textbf{15}, R1233 (2003).

\bibitem{Vlasov-slow-light}
Yu. A. Vlasov and M. O'Boyle and H. F. Hamann and S. J. McNab,
Nature \textbf{438}, 65 (2005).

\bibitem{Fan:2002-908:APL}
S.~H. Fan, Appl. Phys. Lett. {\bf 80},  908  (2002).

\bibitem{Lin:2005-165330:PRB}
L.~L. Lin, Z.~Y. Li, and B. Lin,
Phys. Rev. B {\bf 72}, 165330 (2005).

\bibitem{Kim:2004-5518:OE}
S. Kim, I. Park, H. Lim, and C.~S. Kee, Opt. Express {\bf 12},  5518  (2004).

\bibitem{Wang:2003-66616:PRE}
Z. Wang and S. Fan,
Phys. Rev. E {\bf 68}, 066616 (2003).

\bibitem{Maes:2005-1778:JOSB}
B. Maes, P. Bienstman, and R. Baets,
J. Opt. Soc. Am. B \textbf{19}, 1778 (2005).

\bibitem{Yariv:1999-711:OL}
A. Yariv, Y. Xu, R.~K. Lee, and A. Scherer, Opt. Lett. {\bf 24},
711 (1999).

\bibitem{Johnson:2001-173:OE}
S. Johnson and J.~D. Joannopoulos,
Optics Express \textbf{8}, 173 (2001).

\end{thebibliography}
\end{document}